\documentclass
[aps,pre,twocolumn,groupedaddress,showpacs,amsmath,floatfix]{revtex4}
\usepackage{bm,graphicx}

\begin{document}

\preprint{PUPT-2140}

\title{Field theory of the inverse cascade in two-dimensional turbulence}

\author{Jackson R. Mayo}
\email{jmayo@princeton.edu}
\affiliation{Department of Physics, Princeton University, Princeton, New
Jersey 08544-0708}

\begin{abstract}
A two-dimensional fluid, stirred at high wavenumbers and damped by both
viscosity and linear friction, is modeled by a statistical field theory. The
fluid's long-distance behavior is studied using renormalization-group (RG)
methods, as begun by Forster, Nelson, and Stephen [Phys.\ Rev.\ A
\textbf{16}, 732 (1977)]. With friction, which dissipates energy at low
wavenumbers, one expects a stationary inverse energy cascade for strong
enough stirring. While such developed turbulence is beyond the quantitative
reach of perturbation theory, a combination of exact and perturbative results
suggests a coherent picture of the inverse cascade. The zero-friction
fluctuation-dissipation theorem (FDT) is derived from a generalized
time-reversal symmetry and implies zero anomalous dimension for the velocity
even when friction is present. Thus the Kolmogorov scaling of the inverse
cascade cannot be explained by any RG fixed point. The $\beta$ function for
the dimensionless coupling $\hat g$ is computed through two loops; the $\hat
g^3$ term is positive, as already known, but the $\hat g^5$ term is negative.
An ideal cascade requires a linear $\beta$ function for large $\hat g$,
consistent with a Pad\'e approximant to the Borel transform. The conjecture
that the Kolmogorov spectrum arises from an RG flow through large $\hat g$ is
compatible with other results, but the accurate $k^{-5/3}$ scaling is not
explained and the Kolmogorov constant is not estimated. The lack of scale
invariance should produce intermittency in high-order structure functions, as
observed in some but not all numerical simulations of the inverse cascade.
When analogous RG methods are applied to the one-dimensional Burgers equation
using an FDT-preserving dimensional continuation, equipartition is obtained
instead of a cascade---in agreement with simulations.
\end{abstract}


\maketitle

\section{\label{Intro}Introduction}

The cascade of energy to low wavenumbers in two-dimensional turbulence
\cite{K67}, more than other turbulence problems, is suited to the standard
methods of statistical field theory and the renormalization group (RG). These
methods \cite{Z96}, originating in quantum field theory, show that arbitrary
short-distance interactions lead to long-distance behavior described by a
local, renormalizable action---an effective field theory. Correlation
functions computed from this action contain ultraviolet (UV) divergences that
can be eliminated by redefining the parameters and fields. The divergences
leave their mark, however, in the dependence on the renormalization scale and
the resulting anomalous scaling laws.

The classic application of statistical field theory is to critical phenomena
(second-order phase transitions) in condensed matter \cite{Z96}. The infrared
(IR) scale invariance of correlation functions at the transition temperature
is explained by a fixed point of the RG flow. The inverse energy cascade of
two-dimensional turbulence is likewise believed to be nearly scale invariant,
and one might suspect that a similar fixed point is responsible. We will see,
however, that no fixed point can reproduce the observed $k^{-5/3}$ energy
spectrum. Rather, we will argue that the inverse cascade arises from a
nontrivial RG flow and thus is not expected to be completely scale invariant.

In the study of turbulence, a deviation from scale invariance (a dependence
of dimensionless physical quantities on scale) is referred to as
intermittency \cite{F95}. While intermittency is recognized as a property of
the three-dimensional direct cascade of energy, its existence in the
two-dimensional inverse cascade is unsettled. For a nonstationary inverse
cascade, in which energy is not dissipated but progresses to ever-lower
wavenumbers, intermittency is not observed in numerical simulations
\cite{SY93,SY94}; a theoretical explanation has been given \cite{Y99}. In
this paper we deal solely with the stationary regime, where an inverse
cascade requires a low-wavenumber energy sink. With few exceptions,
simulations of such a cascade confirm the $k^{-5/3}$ energy spectrum
initially predicted \cite{K67} on the basis of scale invariance. But one set
of simulations \cite{BDF95} finds strong intermittency in fourth- and
higher-order velocity correlations. Other simulations \cite{BCV00} and
experiments \cite{PT98}, though, find no significant intermittency. The
various studies differ mainly in the precise form of the dissipation terms.
The evidence suggests that intermittency in the stationary inverse cascade,
permitted by our theory, is at least possibly realized.

It is our restricted focus on the inverse cascade that allows us to work with
a purely local field theory. The random force that stirs the fluid is
correlated over a limited range and is effectively local in a long-distance
description. As with the RG treatment of quantum fields and condensed matter,
we expect all short-distance details to become irrelevant except as they are
manifested in local, renormalizable couplings. The two-dimensional direct
cascade of enstrophy to high wavenumbers \cite{K67} thus falls outside our
scope. A previous RG analysis of two-dimensional turbulence \cite{H98} is
formally similar to ours, but it follows three-dimensional studies by
adopting the long-range force correlations necessary for a direct cascade;
even its derivation of the inverse cascade relies on nonlocal forcing. Here
we apply RG methods in the familiar domain of local field theory, which
should allow a physical treatment of the inverse cascade. The explanation of
the direct enstrophy cascade may rest on entirely different foundations, such
as conformal invariance \cite{P93}.

The work most closely aligned with our theoretical approach is due to
Forster, Nelson, and Stephen (FNS) \cite{FNS77}. At the technical level, our
contribution is to add linear friction to FNS model A in $d = 2$ and compute
the RG flow to the next order of perturbation theory, two loops. The
inclusion of friction, which dissipates energy at low wavenumbers, makes our
theory capable in principle of describing the inverse cascade and its
$k^{-5/3}$ spectrum---unlike FNS model A, which gives a $k^1$ spectrum
corresponding to energy equipartition in $d = 2$. We also note a difference
in our viewpoint from that of FNS and others \cite{DM79,H98,AAV99} who apply
RG methods to turbulence by analogy with critical phenomena. These authors
seek a controlled IR-stable fixed point by starting with a logarithmically
divergent field theory and then decreasing the dimension of space or the
exponent of the stirring-force correlation by $\epsilon$. This adds to the
$\beta$ function a negative linear term proportional to $\epsilon$. With the
usual positive one-loop term, there exists an IR-stable fixed point at a
coupling that goes to zero with $\epsilon$; the fixed-point theory can then
be expanded in $\epsilon$ instead of the original coupling. Like FNS, we work
in $d = 2 - \epsilon$ to regulate UV divergences, but we ultimately take
$\epsilon = 0$, so that the fixed point is trivial. Our inverse-cascade model
lies not at a fixed point but in the region of large dimensionless coupling.

Naturally the use of perturbation theory is questionable for strong coupling.
Our perturbative results will have direct quantitative application only to
the extreme IR limit controlled by the trivial fixed point, which is of some
interest in itself. Nevertheless, we will make reasonable conjectures about
the theory's strong-coupling behavior that are consistent with the inverse
cascade, bearing in mind the dangers of the nonperturbative regime. Besides
the concern with the numerical accuracy of extrapolations, there are
fundamental difficulties at strong coupling. The anomalous dimensions of
operators may be large, and the relevance of terms in the action may differ
from the weak-coupling case. Furthermore, at strong coupling, there is no
simple relation between the couplings in different renormalization schemes,
such as the Wilsonian cutoff (useful for physical interpretation) and minimal
subtraction (convenient for systematic calculation). We may hope that these
subtleties do not affect the main conclusions even at very strong coupling.
At least we know that the theory of critical phenomena in $d = 4 - \epsilon$
is extrapolated to $\epsilon = 1$ (moderate coupling) with acceptable
results.

In Sec.~\ref{Framework} we describe the basis of our theory and our method of
calculation, confirming the one-loop RG flow of FNS \cite{FNS77}. In
Sec.~\ref{Properties} we present symmetries and other properties of the
theory that do not involve a dubious extrapolation to strong coupling. In
Sec.~\ref{TwoLoop} we compute the two-loop term of the $\beta$ function. In
Sec.~\ref{Cascade} we relate the plausible strong-coupling behavior of the
theory to the phenomenology of the inverse cascade. In Sec.~\ref{Burgers}, as
a test of our methods, we consider a rather different model, the UV-stirred
one-dimensional Burgers equation. A summary and discussion are presented in
Sec.~\ref{Discussion}.

\section{\label{Framework}Framework}

\subsection{\label{Path}Path integral for the Navier--Stokes equation}

The Navier--Stokes equation for the velocity field $v_j$ of an incompressible
two-dimensional fluid is
\begin{equation}
\label{NSv}
\dot v_j + v_k \nabla_k v_j + \nabla_j P - \nu\nabla^2 v_j + \alpha v_j =
f_j,
\end{equation}
where $P$ is the pressure divided by the density, $\nu$ is the kinematic
viscosity, $\alpha$ is the friction coefficient, and $f_j$ is the force per
unit mass. The incompressibility condition $\nabla_i v_i = 0$ allows the
velocity to be expressed as
\begin{equation}
v_i = \epsilon_{ij} \nabla_j \psi,
\end{equation}
where $\psi$ is a pseudoscalar field called the stream function and
$\epsilon_{ij}$ is the alternating tensor, which in two dimensions satisfies
\begin{equation}
\epsilon_{ij} \epsilon_{kl} = \delta_{ik} \delta_{jl} - \delta_{il}
\delta_{jk}.
\end{equation}
Upon writing Eq.~(\ref{NSv}) in terms of $\psi$ and applying the operator
$\epsilon_{ij} \nabla_i$, we obtain \cite{H98}
\begin{equation}
\label{NSpsi}
-\nabla^2 \dot\psi - \epsilon_{ij} \nabla_i \nabla^2 \psi \nabla_j \psi +
\nu\nabla^4 \psi - \alpha\nabla^2 \psi = \eta.
\end{equation}
Here
\begin{equation}
\eta = \bm{\nabla} \times \mathbf{f},
\end{equation}
with the notation
\begin{equation}
\mathbf{a} \times \mathbf{b} \equiv \epsilon_{ij} a_i b_j.
\end{equation}

In the real three-dimensional world, Eq.~(\ref{NSv}) is a good approximation
for a thin fluid film provided either (a) the boundary surface(s) and the
coordinate system are rotating rapidly about a perpendicular axis or (b) the
fluid is conducting and subject to a strong perpendicular magnetic field
\cite{L97}. In either case, boundary-layer effects produce a linear friction
parametrized by $\alpha$, which has units of frequency.

For convenience, our field-theory calculations will use the method of
dimensional regularization \cite{TV72}, based on continuation to a noninteger
spatial dimension $d = 2 - \epsilon$. Because in the end we are concerned
only with $d = 2$, we adopt a formal continuation of the Navier--Stokes
equation that preserves its two-dimensional features. For general $d$, we
retain the stream-function representation by choosing a ``physical''
two-dimensional subspace that contains the tensor $\epsilon_{ij}$ and the
external wavevectors of correlation functions. Denoting by $\Theta_{ij}$ the
projector onto the physical subspace, we now have
\begin{equation}
\label{epseps}
\epsilon_{ij} \epsilon_{kl} = \Theta_{ik} \Theta_{jl} - \Theta_{il}
\Theta_{jk}.
\end{equation}
We take Eq.~(\ref{NSpsi}) as the equation of motion for $\psi$, with
$\nabla^2 \equiv \nabla_k \nabla_k$ interpreted as the $d$-dimensional
Laplacian. In this way we preserve (for $\nu=\alpha=\eta=0$) the formal
conservation of the energy and enstrophy,
\begin{align}
E &= \tfrac{1}{2} \int d^d x\, \nabla_i \psi \nabla_i \psi,\\
\mathnormal{\Omega} &= \tfrac{1}{2} \int d^d x\, \nabla^2 \psi \nabla^2 \psi.
\end{align}

We assume that the fluid is stirred by a Gaussian random force that is
uncorrelated in time \cite{FNS77}, with
\begin{align}
\langle f_i(\omega,\mathbf{k})\, f_j(\omega',\mathbf{k}')\rangle =
{}&(2\pi)^{d+1} \delta(\omega+\omega')\,
\delta(\mathbf{k}+\mathbf{k}')\nonumber\\
\label{forcing}
&\times (\delta_{ij} - k_i k_j/k^2)\, D(k^2),
\end{align}
so that $\eta$ is also Gaussian with
\begin{equation}
\langle\eta(\omega,\mathbf{k})\, \eta(\omega',\mathbf{k}')\rangle =
(2\pi)^{d+1} \delta(\omega+\omega')\, \delta(\mathbf{k}+\mathbf{k}')\, k^2
D(k^2).
\end{equation}
A classical system with random forcing can be treated in the formalism of
quantum field theory \cite{MSR73}, including the path-integral representation
\cite{BJW76,P77,AVP83}. Upon introduction of a pseudoscalar field $p$
conjugate to $\eta$, correlation functions of $\psi$ are given by the path
integral
\begin{equation}
\langle F[\psi]\rangle \propto \int \mathcal{D}\psi\, \mathcal{D}p\,
F[\psi]\, e^{-S},
\end{equation}
with the action
\begin{align}
S = {}&{\int dt}\, d^d x\, \bigl[\tfrac{1}{2} (-\nabla^2 p)\,
D(-\nabla^2)p\nonumber\\
&+ ip(-\nabla^2 \dot\psi - \epsilon_{ij} \nabla_i \nabla^2 \psi \nabla_j \psi
+ \nu\nabla^4 \psi - \alpha\nabla^2 \psi)\bigr]\nonumber\\
= {}&{\int dt}\, d^d x\, \bigl[\tfrac{1}{2} (-\nabla^2 p)\, D(-\nabla^2)p +
i\nu p\nabla^4 \psi\nonumber\\
\label{Sppsi}
&- i\alpha p\nabla^2 \psi - ip\nabla^2 \dot\psi - i\epsilon_{ij} \nabla_i
\nabla_k p \nabla_j \psi \nabla_k \psi\bigr],
\end{align}
where we have integrated the $p\psi\psi$ term by parts. The Jacobian
determinant from changing variables from $\eta$ to $\psi$ is an unimportant
constant by virtue of causality \cite{AVP83}.

\subsection{\label{Relevance}Relevance of couplings}

In the field theory based on the action (\ref{Sppsi}), the long-distance
behavior is governed by just the renormalizable terms---those with
coefficients whose scaling dimensions with respect to wavenumber in $d = 2$
are zero (marginal) or positive (relevant) \cite{Z96}. We assign scaling
dimensions $d_\psi$, $d_p$, $d_t$ to the fields $\psi$ and $p$ and to the
time $t$ by requiring that the highest-derivative quadratic terms in the
action have dimensionless coefficients in $d = 2$, since these terms control
the asymptotic behavior of propagators and thus the UV convergence or
divergence of diagrams. Because of the additional derivatives, the viscous
term $i\nu p\nabla^4 \psi$ is clearly less relevant than the friction term
$-i\alpha p\nabla^2 \psi$. In fact, it is commonly said that viscosity is
irrelevant to the inverse cascade, but we now show that this cannot be taken
in the technical sense. If the viscous term is ignored, then the $p\nabla^2
\psi$ and $p\nabla^2 \dot\psi$ terms have dimensionless coefficients only if
$d_t = 0$ and $d_\psi = -d_p$. For a nontrivial theory, the $p\psi\psi$ term
must be renormalizable, giving $2 \ge 4 + d_p + 2d_\psi = 4 - d_p$. With $d_p
\ge 2$, there exists no local renormalizable forcing term quadratic in $p$.

Let us therefore retain the viscous term and recompute the scaling
dimensions. The $p\nabla^4 \psi$ and $p\nabla^2 \dot\psi$ terms possess
dimensionless coefficients only if $d_t = -2$ and $d_\psi = -d_p$.
Renormalizability of the $p\psi\psi$ term now gives $4 \ge 4 + d_p + 2d_\psi
= 4 - d_p$, so that $d_p \ge 0$. It remains to specify the forcing term. We
assume that the external forcing is confined to a band of high wavenumbers,
as in model C of FNS \cite{FNS77}. The effective forcing at low wavenumbers
is generated by renormalization; because the interaction in Eq.~(\ref{Sppsi})
has two spatial derivatives acting on $p$, at least two derivatives must
accompany each factor of $p$ in any term so generated. The only
renormalizable forcing term is then $\frac{1}{2} D_0 \nabla^2 p \nabla^2 p$,
whose coefficient is dimensionless for $d_p = d_\psi = 0$. This effective
forcing has $D(k^2) = D_0 k^2$, as in model A of FNS. All terms in the action
are now marginal, except for the friction term, which is relevant
(coefficient of dimension $2$). The only other renormalizable terms that
could be generated are ones containing only $\psi$, but these are not
generated (see Sec.~\ref{Feynman}).

Next we label the fields and the time in Eq.~(\ref{Sppsi}) with the subscript
``phys'' and introduce rescaled variables to simplify the action. Tentatively
seeking to set all coefficients other than forcing and friction equal to $1$,
we take
\begin{equation}
\begin{aligned}
\label{altpsipt}
\psi_{\text{phys}} &= \nu\psi, & p_{\text{phys}} &= \nu^{-1} p, &
t_{\text{phys}} &= \nu^{-1} t.
\end{aligned}
\end{equation}
The result is
\begin{align}
S = {}&{\int dt}\, d^d x\, \bigl(\tfrac{1}{2} g^2 \nabla^2 p \nabla^2 p +
ip\nabla^4 \psi\nonumber\\
\label{Sippsi}
&- i\nu^{-1} \alpha p\nabla^2 \psi - ip\nabla^2 \dot\psi - i\epsilon_{ij}
\nabla_i \nabla_k p \nabla_j \psi \nabla_k \psi\bigr),
\end{align}
where
\begin{equation}
g = D_0^{1/2} \nu^{-3/2}.
\end{equation}
We adopt, however, a different rescaling that will be particularly convenient
in Sec.~\ref{FDT}:
\begin{equation}
\begin{aligned}
\label{ppsit}
\psi_{\text{phys}} &= g\nu\psi, & p_{\text{phys}} &= (g\nu)^{-1} p, &
t_{\text{phys}} &= (g\nu)^{-1} t.
\end{aligned}
\end{equation}
The final form of the action is then
\begin{align}
S = {}&{\int dt}\, d^d x\, \bigl(\tfrac{1}{2} g^{-1} \nabla^2 p \nabla^2 p +
ig^{-1} p\nabla^4 \psi\nonumber\\
\label{Sfppsi}
&- i\rho p\nabla^2 \psi - ip\nabla^2 \dot\psi - i\epsilon_{ij} \nabla_i
\nabla_k p \nabla_j \psi \nabla_k \psi\bigr),
\end{align}
where
\begin{equation}
\rho = (g\nu)^{-1} \alpha = D_0^{-1/2} \nu^{1/2} \alpha.
\end{equation}
For general $d = 2 - \epsilon$, the scaling dimensions in Eq.~(\ref{Sfppsi})
are
\begin{equation}
\label{deps}
\begin{aligned}
d_\psi = d_p &= -\tfrac{1}{2} \epsilon, & d_t &= -2 + \tfrac{1}{2}
\epsilon,\\
d_g &= +\tfrac{1}{2} \epsilon, & d_\rho &= +2 - \tfrac{1}{2} \epsilon.
\end{aligned}
\end{equation}
In two dimensions, $g$ is a dimensionless coupling and $\rho$ is analogous to
a mass parameter in quantum field theory.

\subsection{\label{Feynman}Feynman rules}

\begin{figure}
\includegraphics{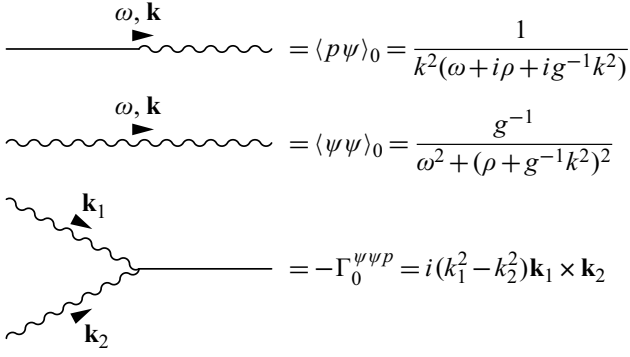}%
\caption{\label{FeynRules}Feynman rules for the action
(\protect\ref{Sfppsi}).}
\end{figure}

Correlation functions can be calculated for the action (\ref{Sfppsi}) using
Feynman diagrams whose lines carry both frequencies and wavevectors. We
represent the fields $\psi$ and $p$ by wiggly and plain lines respectively.
The ingredients of the diagrams, shown in Fig.~\ref{FeynRules}, are the
propagators $\langle p\psi\rangle_0$, $\langle\psi\psi\rangle_0$, obtained
from the quadratic terms in the action, and the vertex factor
$-\Gamma^{\psi\psi p}_0$, obtained from the cubic term. We label these
quantities with the subscript $0$ because they are the tree-level
contributions to the exact two-point correlation functions $\langle
p\psi\rangle$, $\langle\psi\psi\rangle$ and the exact three-point
one-particle-irreducible (1PI) function $-\Gamma^{\psi\psi p}$.

The remaining contributions arise from diagrams containing loops. For these
diagrams, we integrate over each loop frequency and wavevector according to
\begin{equation}
\int \frac{d\omega}{2\pi} \frac{d^d k}{(2\pi)^d}.
\end{equation}
Because the integrand is a rational function of the frequencies, the $\omega$
integrations can easily be performed by the contour method before integrating
over wavevectors. This contour integration shows that a 1PI diagram (or
subdiagram) vanishes if all its external lines are $\psi$ \cite{AVP83}, since
there is a closed loop of $\langle p\psi\rangle_0$ propagators and the
integrand is an analytic function of the loop frequency in the upper
half-plane.

Unlike many field theories in a low number of spatial dimensions, ours does
not contain IR divergences even for $\rho = 0$. This is because, after
integration over frequencies, internal-line propagators scale as $k^{-2}$,
but at least one further factor of the wavevector arises from each of the two
vertices that a line connects. Hence the integrand does not diverge as the
wavevector of any internal line goes to zero. The frequencies and wavevectors
of the external lines act as an IR cutoff. For simplicity, our calculations
will adopt another IR cutoff by assuming that $\rho > 0$; then 1PI diagrams
are analytic at zero external frequencies and wavevectors.

\subsection{\label{OneLoop}One-loop renormalization}

\begin{figure}
\includegraphics{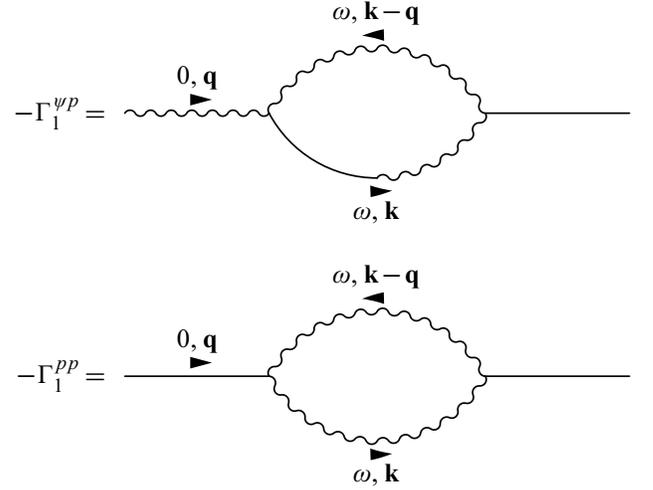}%
\caption{\label{1L1PI}One-loop diagrams for the two-point 1PI functions
$-\Gamma^{\psi p}$ and $-\Gamma^{pp}$ at zero frequency.}
\end{figure}

The coefficients of the quadratic terms in the action (with frequency $\chi$
and wavevector $\mathbf{q}$) are
\begin{align}
\Gamma^{\psi p}_0 &= q^2 (\chi + i\rho + ig^{-1} q^2),\\
\Gamma^{pp}_0 &= g^{-1} q^4.
\end{align}
These are corrected at one loop by the two-point 1PI diagrams in
Fig.~\ref{1L1PI}. Because of the external wavevectors in the vertex factors,
the diagrams are $O(q^4)$, and so we can set $\chi = 0$. We now demonstrate
the computation of the $-\Gamma^{\psi p}_1$ diagram, to show the basic
methods to be used for two-loop diagrams in Sec.~\ref{TwoLoop}.

The frequency integral of the propagators is
\begin{align}
I &\equiv \int_{-\infty}^\infty \frac{d\omega}{2\pi} \bigl[g^{-1} k^{-2}
(\omega + i\rho + ig^{-1} k^2)^{-1}\nonumber\\
&\times (\omega + i\rho + ig^{-1} |\mathbf{k}-\mathbf{q}|^2)^{-1} (\omega -
i\rho - ig^{-1} |\mathbf{k}-\mathbf{q}|^2)^{-1}\bigr]\nonumber\\
&= \frac{-ig/4k^2}{(g\rho + k^2 - \mathbf{k}\cdot\mathbf{q} + \frac{1}{2}q^2)
(g\rho + k^2 - 2\mathbf{k}\cdot\mathbf{q} + q^2)},
\end{align}
obtained conveniently by closing the contour in the upper half-plane and
picking up one pole. We next multiply by the vertex factors and expand to
$O(q^4)$:
\begin{align}
i(k^2 - 2\mathbf{k}&\cdot\mathbf{q}) (\mathbf{k}\times\mathbf{q})\,
i(2\mathbf{k}\cdot\mathbf{q} - q^2) (\mathbf{k}\times\mathbf{q})\,
I\nonumber\\
= {}&(k^2 - 2\mathbf{k}\cdot\mathbf{q}) (q^2 - 2\mathbf{k}\cdot\mathbf{q})
[k^2_\parallel q^2 - (\mathbf{k}\cdot\mathbf{q})^2] I\nonumber\\
\to {}&\frac{6ig (\mathbf{k}\cdot\mathbf{q})^2 [k^2_\parallel q^2 -
(\mathbf{k}\cdot\mathbf{q})^2]} {4(g\rho + k^2)^3}\nonumber\\
\label{q4}
&- \frac{ig [k^2 q^2 + 4(\mathbf{k}\cdot\mathbf{q})^2] [k^2_\parallel q^2 -
(\mathbf{k}\cdot\mathbf{q})^2]}{4k^2 (g\rho + k^2)^2},
\end{align}
where $k^2_\parallel \equiv \Theta_{ij} k_i k_j$ is the squared projection of
$\mathbf{k}$ onto the physical subspace and we infer from Eq.~(\ref{epseps})
that
\begin{equation}
(\mathbf{k} \times \mathbf{q})^2 = k^2_\parallel q^2 - (\mathbf{k} \cdot
\mathbf{q})^2.
\end{equation}
We omit $O(q^3)$ terms in Eq.~(\ref{q4}) because they will now disappear when
we average over directions of $\mathbf{k}$.

With the $d$-dimensional isotropization formulas
\begin{equation}
\begin{aligned}
k_i k_j &\to k^2 \frac{\delta_{ij}}{d}, & k_i k_j k_k k_l &\to k^4
\frac{\delta_{ij}\delta_{kl} + \delta_{ik}\delta_{jl} +
\delta_{il}\delta_{jk}}{d(d+2)},
\end{aligned}
\end{equation}
which imply
\begin{equation}
\begin{aligned}
\label{isotrop}
k^2_\parallel &\to \frac{2k^2}{d}, & (\mathbf{k}\cdot\mathbf{q})^2 &\to
\frac{k^2 q^2}{d},\\
(\mathbf{k}\cdot\mathbf{q})^2 k^2_\parallel &\to \frac{4k^4 q^2}{d(d + 2)}, &
(\mathbf{k}\cdot\mathbf{q})^4 &\to \frac{3k^4 q^4}{d(d + 2)},
\end{aligned}
\end{equation}
the value of the diagram to $O(q^4)$ becomes
\begin{align}
-\Gamma^{\psi p}_1 = {}&{-}igq^4\nonumber\\
&\times \int_0^\infty \frac{dk}{(2\pi)^d}
\frac{2\pi^{d/2}k^{d-1}}{\Gamma(\frac{1}{2}d)} k^2 \frac{(6 + d) g\rho +
dk^2} {4d(d + 2) (g\rho + k^2)^3}\nonumber\\
= {}&{-}igq^4\frac{6 - d}{(2 + d) 2^{4+d}\, \Gamma(\frac{1}{2}d)
\sin(\frac{1}{2}\pi d)} \left(\frac{\pi}{g\rho}\right)^{1-d/2}.
\end{align}
For $d = 2 - \epsilon$ with $\epsilon \to 0$, we find
\begin{equation}
\label{1Lpsip}
+\Gamma^{\psi p}_1 = igq^4 \left(\frac{1}{32\pi\epsilon} -
\frac{\ln(g\rho/4\pi) + \gamma_{\text{E}} - 1}{64\pi} + O(\epsilon)\right),
\end{equation}
where $\gamma_{\text{E}}$ is the Euler constant. The result for the other
one-loop diagram (taking into account the symmetry factor) is very similar:
\begin{equation}
\label{1Lpp}
+\Gamma^{pp}_1 = gq^4 \left(\frac{1}{32\pi\epsilon} - \frac{\ln(g\rho/4\pi) +
\gamma_{\text{E}}}{64\pi} + O(\epsilon)\right).
\end{equation}
We have carefully obtained the $O(\epsilon^0)$ terms, which will be important
in Sec.~\ref{TwoLoop}.

The method of minimal subtraction \cite{T73} expresses the bare couplings $g$
and $\rho$, of respective dimensions $\frac{1}{2}\epsilon$ and $2 -
\frac{1}{2}\epsilon$, in terms of a renormalization scale $\mu$ (dimension
$1$) and dimensionless renormalized couplings $\bar g$ and $\bar\rho$ as
\begin{align}
g &= \mu^{\epsilon/2} \bigl[\bar g + \bar g^3 a_1 \epsilon^{-1} + O(\bar
g^5)\bigr],\\
\rho &= \mu^{2-\epsilon/2} \bar\rho,
\end{align}
where the corrections (called counterterms) involve only negative powers of
$\epsilon$. The coefficient of the $O(\bar g^3)$ counterterm, and the absence
of any counterterms for $\bar\rho$, are obtained by requiring that
$\Gamma^{\psi p}$ and $\Gamma^{pp}$ to $O(q^4)$ be finite at $\epsilon = 0$
when expressed in terms of $\bar g$ and $\bar\rho$:
\begin{alignat}{2}
\Gamma^{\psi p} = {}&q^2 (\chi + i\rho)\span\span\nonumber\\
&+ iq^4 \left[g^{-1} + g \left(\frac{1}{32\pi\epsilon} + O(\epsilon^0)\right)
+ O(g^3)\right]\span\span\nonumber\\
= {}&q^2 (\chi + i\mu^{2-\epsilon/2} \bar\rho)\span\span\nonumber\\
&+ {}& &iq^4 \mu^{-\epsilon/2}\nonumber\\
&&&\times \left[\bar g^{-1} + \bar g \left(\frac{1}{32\pi\epsilon} -
\frac{a_1}{\epsilon} + O(\epsilon^0)\right) + O(\bar g^3)\right],
\displaybreak[0]\\
\Gamma^{pp} = {}&q^4 \mu^{-\epsilon/2}\span\span\nonumber\\
&\times \left[\bar g^{-1} + \bar g \left(\frac{1}{32\pi\epsilon} -
\frac{a_1}{\epsilon} + O(\epsilon^0)\right) + O(\bar g^3)\right].\span\span
\end{alignat}
Hence
\begin{equation}
a_1 = \frac{1}{32\pi},
\end{equation}
and no counterterms of any order are needed for $\bar\rho$ because the $q^2$
term of $\Gamma^{\psi p}$ is not renormalized.

As for the three-point 1PI function $-\Gamma^{\psi\psi p}$, we will see in
Sec.~\ref{Galilean} that it is related by Galilean invariance to the
$q^2\chi$ term of $\Gamma^{\psi p}$. Because the latter term is not
renormalized, we have
\begin{equation}
-\Gamma^{\psi\psi p} = -\Gamma^{\psi\psi p}_0 = i(k_1^2 - k_2^2) \mathbf{k}_1
\times \mathbf{k}_2,
\end{equation}
up to irrelevant terms with more factors of wavevector. We have thus rendered
the theory finite at one loop by renormalizing only the coupling $\bar g$,
without the need for counterterms to rescale the fields $\psi$ and $p$ or the
time $t$. Crucial for this was the equality of the $1/32\pi\epsilon$ terms in
Eqs.\ (\ref{1Lpsip}) and (\ref{1Lpp}). We conclude that the anomalous
dimensions are zero at one loop, and in Sec.~\ref{FDT} we will show that in
minimal subtraction they are exactly zero,
\begin{equation}
\gamma_\psi = \gamma_p = \gamma_t = 0.
\end{equation}
In a different renormalization scheme, or with a different definition of the
fields and the time such as Eq.~(\ref{altpsipt}), the anomalous dimensions
would not vanish identically, but at any RG fixed point their values are
universal \cite{Z96} and so they would be zero there.

The $\beta$ functions for the dimensionless couplings are determined by the
RG invariance of the bare couplings,
\begin{equation}
\left(\mu \frac{\partial}{\partial\mu} + \beta(\bar g)
\frac{\partial}{\partial\bar g} + \beta(\bar\rho)
\frac{\partial}{\partial\bar\rho}\right) \begin{Bmatrix} g\\ \rho
\end{Bmatrix} = 0.
\end{equation}
We obtain
\begin{align}
\beta(\bar g) &= -\tfrac{1}{2} \epsilon \bar g + \frac{\bar g^3}{32\pi} +
O(\bar g^5),\\
\beta(\bar\rho) &= \bigl(-2 + \tfrac{1}{2} \epsilon\bigr) \bar\rho.
\end{align}
The force correlation adopted by FNS \cite{FNS77} differs from
Eq.~(\ref{forcing}) by a factor of $2$, and consequently the dimensionless
coupling of FNS is $\bar\lambda = 2^{-1/2} \bar g$. Thus we have confirmed
the FNS result
\begin{equation}
\beta(\bar\lambda) = -\tfrac{1}{2} \epsilon \bar\lambda +
\frac{\bar\lambda^3}{16\pi} + O(\bar\lambda^5).
\end{equation}

\section{\label{Properties}General properties}

\subsection{\label{Galilean}Galilean invariance}

In two dimensions, the equation of motion (\ref{NSpsi}) and thus the action
(\ref{Sfppsi}) have the important physical property of invariance under a
Galilean transformation to a reference frame moving with constant velocity
$\mathbf{u}$ \cite{FNS77,DM79,AVP83}. This property depends on the assumption
that the stirring force is uncorrelated in time, since otherwise there would
exist a link between a point in space at one time and a ``corresponding''
point in space at a different time. Specifically, the action (\ref{Sfppsi})
is invariant under the transformation
\begin{equation}
\begin{aligned}
\label{GalTrans}
\psi(t, \mathbf{x}) &\to \psi(t, \mathbf{x} + \mathbf{u}t) + \epsilon_{ij}
x_i u_j,\\
p(t, \mathbf{x}) &\to p(t,\mathbf{x} + \mathbf{u}t),
\end{aligned}
\end{equation}
which induces the familiar transformation of the velocity,
\begin{equation}
\mathbf{v}(t, \mathbf{x}) \to \mathbf{v}(t, \mathbf{x} + \mathbf{u}t) -
\mathbf{u}.
\end{equation}
The original Navier--Stokes equation (\ref{NSv}) is not Galilean invariant
because of the friction term, which introduces a preferred state of rest; but
the differentiation in deriving the stream-function equation (\ref{NSpsi})
eliminates the constant shift in $\mathbf{v}$ \cite{H98}.

We would not expect Galilean invariance in $d = 2$ to be preserved by our
renormalization method unless the theory remains Galilean invariant when
dimensionally regulated. We now show that our formal stream-function
representation in arbitrary $d$ is invariant under the transformation
(\ref{GalTrans}), provided that $\mathbf{u}$ lies in the physical subspace.
For convenience we use the infinitesimal form
\begin{equation}
\begin{aligned}
\label{InfGalTrans}
\delta\psi &= t u_i \nabla_i \psi + \epsilon_{ij} x_i u_j, & \delta p &= t
u_i \nabla_i p.
\end{aligned}
\end{equation}
In the corresponding variation of the action (\ref{Sfppsi}), terms
proportional to $t$ vanish automatically because they correspond to a simple
spatial translation. The interesting terms are those where the $t$ is
differentiated ($-ip \nabla^2 \dot\psi$) and where $\psi$ is varied by
$\epsilon_{ij} x_i u_j$. Neither of these affects the forcing, viscous, or
friction terms, which contain $\psi$ only as $\nabla^2 \psi$ and are
trivially invariant. We are left with
\begin{align}
\delta S = -\int dt\, d^d x\, (&ipu_i \nabla_i \nabla^2 \psi + i\epsilon_{ij}
\nabla_i \nabla_k p\, \epsilon_{jl} u_l \nabla_k \psi\nonumber\\
&+ i\epsilon_{ij} \nabla_i \nabla_k p \nabla_j \psi\, \epsilon_{kl} u_l).
\end{align}
Upon integration by parts and use of $\Theta_{il} u_l = u_i$, the first two
terms cancel and the third vanishes.

In Fourier space, the transformation (\ref{InfGalTrans}) becomes
\begin{align}
\delta\psi(\omega,\mathbf{k}) &= \mathbf{u} \cdot \mathbf{k}\,
\frac{\partial}{\partial\omega} \psi(\omega,\mathbf{k}) - i(2\pi)^{d+1}
\delta(\omega)\, \mathbf{u} \times \bm{\nabla} \delta(\mathbf{k}),\nonumber\\
\delta p(\omega,\mathbf{k}) &= \mathbf{u} \cdot \mathbf{k}\,
\frac{\partial}{\partial\omega} p(\omega,\mathbf{k}).
\end{align}
The action is Galilean invariant by virtue of the relation
\begin{equation}
\mathbf{u} \cdot \mathbf{k}\, \frac{\partial}{\partial\omega} \Gamma^{\psi
p}_0(\omega,\mathbf{k}) = i \mathbf{u} \times \bm{\nabla}' \Gamma^{\psi\psi
p}_0(\omega,\mathbf{k}; 0,\mathbf{k}') |_{\mathbf{k}'=\mathbf{0}};
\end{equation}
both sides equal $\mathbf{u} \cdot \mathbf{k}\, k^2$. The same relation must
then hold between the exact 1PI functions: The coefficient of $k^2 \omega$ in
$\Gamma^{\psi p}$ equals the coefficient of $-i(k_1^2 - k_2^2) \mathbf{k}_1
\times \mathbf{k}_2$ in $\Gamma^{\psi\psi p}$. Since the former is not
renormalized, neither is the latter.

\subsection{\label{FDT}Fluctuation-dissipation theorem}

For zero friction ($\rho = 0$), the action (\ref{Sfppsi}) is equivalent to
model A of FNS \cite{FNS77}, who note that it obeys detailed balance and thus
is subject to a classical fluctuation-dissipation theorem (FDT) \cite{DH75}.
A complicated diagrammatic argument demonstrates that the FDT is preserved to
all orders of renormalization \cite{DH75}. Here we reach this conclusion by
obtaining the FDT from an exact symmetry of the action in arbitrary $d$.
Under the formal discrete transformation
\begin{equation}
p \to p - 2i\psi,
\end{equation}
the action with $\rho = 0$ changes only by reversing the sign of the viscous
term. The change in the interaction term vanishes upon integration by parts,
just as in deriving conservation of energy. To restore the sign of the
viscous term, we further perform a complete time reversal,
\begin{equation}
\begin{aligned}
t &\to -t, & \psi &\to -\psi,
\end{aligned}
\end{equation}
which naturally reverses the sign of the dissipation. The net effect is the
transformation
\begin{equation}
\begin{aligned}
p(t) &\to p(-t) + 2i\psi(-t), & \psi(t) &\to -\psi(-t),
\end{aligned}
\end{equation}
a generalized time reversal that is its own inverse and leaves the action
invariant.

The FDT is derived from this symmetry by expressing the invariance of
$\langle p\psi\rangle$:
\begin{equation}
\label{FDTeq}
\langle p\psi\rangle = -\langle\psi p\rangle - 2i \langle\psi\psi\rangle.
\end{equation}
By invariance under time translations and spatial rotations, negating the
times in a two-point correlation function is equivalent to interchanging the
points. As a first application of the FDT, we make use of the theorem that
the equal-time correlation function $\langle p\psi\rangle_= = \langle\psi
p\rangle_=$ is exact at tree level \cite{AVP83}. Thus, for $\rho = 0$, the
exact equal-time stream-function correlation is
\begin{align}
\langle\psi\psi\rangle_= &= \tfrac{1}{2}i \bigl(\langle p\psi\rangle_= +
\langle\psi p\rangle_=\bigr)\nonumber\\
&= \tfrac{1}{2}i \int_{-\infty}^\infty \frac{d\omega}{2\pi} \bigl(\langle
p\psi\rangle_0 + \langle\psi p\rangle_0\bigr)\nonumber\\
&= \tfrac{1}{2}i \int_{-\infty}^\infty \frac{d\omega}{2\pi}
\frac{-2ig^{-1}}{\omega^2 + g^{-2} k^4} = \frac{1}{2k^2}.
\end{align}
The energy spectrum, in the units implied by Eq.~(\ref{ppsit}), is then
\begin{equation}
E(k) = \frac{\pi k}{(2\pi)^2} k^2 \langle\psi\psi\rangle_= = \frac{k}{8\pi}.
\end{equation}
This is an equipartition spectrum, in agreement with FNS model A
\cite{FNS77}.

For $\rho = 0$, the FDT (\ref{FDTeq}) also implies that, if $\langle
p\psi\rangle$ and thus $\langle\psi p\rangle$ are made finite by
renormalizing $\bar g$, then $\langle\psi\psi\rangle$ is likewise finite,
without the need for field or time rescalings. As we have seen, the
three-point 1PI function is automatically finite. Hence the anomalous
dimensions vanish exactly for $\rho = 0$. But in minimal subtraction, the
counterterms for rescalings and for dimensionless couplings are independent
of the mass parameter \cite{T73}. We conclude that in minimal subtraction,
even with friction,
\begin{equation}
\gamma_\psi = \gamma_p = \gamma_t = 0,
\end{equation}
and $\beta(\bar g)$ depends only on $\bar g$. Indeed, we have seen these
statements verified to one loop in Sec.~\ref{OneLoop}.

\subsection{\label{RG}Renormalization-group flows}

We have shown that the anomalous dimensions vanish, and that the renormalized
couplings in exactly two dimensions obey
\begin{align}
\label{RGgbar}
\mu \frac{d\bar g}{d\mu} &\equiv \beta(\bar g) = \frac{\bar g^3}{32\pi} +
O(\bar g^5),\\
\label{RGrhobar}
\bar\rho &= \mu^{-2} \rho.
\end{align}
Thus $\bar\rho$ is very simply related to $\mu$ and can be used to
parametrize it. In the RG flow to low wavenumbers, $\bar\rho$ steadily
increases as friction becomes more important. Meanwhile, $\bar g$ flows in
its own characteristic way regardless of the value of $\bar\rho$; the most we
can say is that once $\bar g$ becomes small, it decreases further and
further, approaching zero in the IR limit. The solution of Eq.~(\ref{RGgbar})
in this limit is
\begin{equation}
\begin{aligned}
\bar g(\mu) &= \sqrt{\frac{16\pi}{\ln(k_g/\mu)}} & (\mu &\ll k_g),
\end{aligned}
\end{equation}
where $k_g$ is the scale at which $\bar g$ becomes large.

In the IR limit, we expect good accuracy from perturbative results such as
the tree-level expression for the energy spectrum,
\begin{equation}
\label{Ekp}
E(k) = \frac{\pi k}{(2\pi)^2} k^2 \langle\psi\psi\rangle_= =
\frac{k^3}{8\pi(g\rho + k^2)}.
\end{equation}
We might suppose that the true asymptotic behavior is given by replacing $g$
with the renormalized coupling $\bar g(k)$. This would follow from RG theory
if the loop corrections to Eq.~(\ref{Ekp}) contained $\ln(k^2)$. But with
$g\rho$ providing an IR cutoff, the diagrams are regular as $k \to 0$ and
instead contain $\ln(g\rho)$. Hence, for the purposes of correlation
functions, the RG flow effectively halts for wavenumbers below the ``mass''
$\sqrt{g\rho}$. If $m$ is the suitably renormalized value of this mass, then
as $k \to 0$ we expect that
\begin{equation}
\label{Ekp1}
E(k) = \frac{k^3}{8\pi\rho\, \bar g(m)}.
\end{equation}
On the other hand, in the bare tree-level result (\ref{Ekp}), $g$ can be
interpreted as a coupling renormalized at a very high wavenumber (the forcing
scale or UV cutoff). With $\beta(\bar g) > 0$, we have $\bar g(m) < g$, and
so the RG result (\ref{Ekp1}) gives a greater $E(k)$ at low $k$. Whereas the
bare tree-level calculation ignores all interactions between scales, the
effect of renormalization is to place more energy and dissipation at low $k$
(and therefore less at high $k$), consistent with the inverse cascade.

From Eq.~(\ref{RGrhobar}), any RG fixed point $(\bar g_*,\bar\rho_*)$ must
have $\bar\rho_* = 0$. Not only do the anomalous dimensions vanish at any
fixed point, but for $\bar\rho = 0$ we have the exact equipartition spectrum
$E(k) \propto k^1$, whether $\bar g$ is at a fixed point or not. It is clear
that, despite the suggestive evidence of scale invariance of the inverse
cascade, an RG fixed point in our framework cannot be the explanation for the
observed $k^{-5/3}$ spectrum. Nevertheless our theory contains all the
essential ingredients that have produced the stationary inverse cascade
experimentally and numerically. A natural explanation is that the $k^{-5/3}$
spectrum arises from the nonperturbative behavior of correlation functions at
$\bar\rho > 0$ and at large values of $\bar g$ that flow rapidly with scale.
Although it is far from obvious how such an RG flow can produce approximate
scale invariance with an effective anomalous dimension, we are motivated to
seek hints about the theory's strong-coupling behavior. The first step, which
can be useful for more mundane purposes as well, is to extend the
renormalization to the next order of perturbation theory.

\section{\label{TwoLoop}Two-loop renormalization}

\subsection{\label{Prescrip}Renormalization prescription and diagrams}

Calculating $\beta(\bar g)$ consistently to two loops requires a precise
specification of the renormalization scheme. Remarkably, though, as long as
$\beta(\bar g)$ depends only on $\bar g$, the $\beta$ function to two loops
(but no further) is independent of the particular scheme chosen \cite{Z96}.
As in Sec.~\ref{OneLoop}, we take $\rho > 0$, compute two-point 1PI diagrams
to $O(q^4)$ expanded about $\epsilon = 0$, and renormalize by minimal
subtraction. The two-loop expression for the bare coupling in terms of the
renormalized coupling is
\begin{equation}
\label{2Lg}
g = \mu^{\epsilon/2} \bigl[\bar g + \bar g^3 a_1 \epsilon^{-1} + \bar g^5
(a_2 \epsilon^{-2} + a_2' \epsilon^{-1}) + O(\bar g^7)\bigr].
\end{equation}
With zero anomalous dimensions, $\beta(\bar g)$ is the only RG function to be
determined and we need only compute a single 1PI function to two loops. Below
we will choose $-\Gamma^{pp}$ because its two-loop diagrams are technically
simpler than those of $-\Gamma^{\psi p}$.

We have seen that $g\rho$, which has dimension $2$ independent of $\epsilon$,
acts as the IR cutoff for wavevector integrals. By dimensional analysis,
$\Gamma^{pp}$ to $O(q^4)$ has the form
\begin{alignat}{2}
\Gamma^{pp} &= q^4 &&\biggl[g^{-1} + g (g\rho)^{-\epsilon/2}
\left(\frac{b_1}{\epsilon} + b_1' + O(\epsilon)\right)\nonumber\\
&&&+ g^3 (g\rho)^{-\epsilon} \left(\frac{b_2}{\epsilon^2} +
\frac{b_2'}{\epsilon} + O(\epsilon^0)\right) + O(g^5)\biggr]\nonumber\\
&= q^4 &&\biggl[g^{-1} + g \left(\frac{b_1}{\epsilon} + b_1' - \tfrac{1}{2}
b_1 \ln(g\rho) + O(\epsilon)\right)\nonumber\\
\label{2Lpp}
&&&+ g^3 \left(\frac{b_2}{\epsilon^2} + \frac{b_2' - b_2
\ln(g\rho)}{\epsilon} + O(\epsilon^0)\right) + O(g^5)\biggr].
\end{alignat}
To simplify the two-loop calculations we formally set $\rho = g^{-1}$ and
restore the $\ln(g\rho)$ terms at the end.

\begin{figure}
\includegraphics{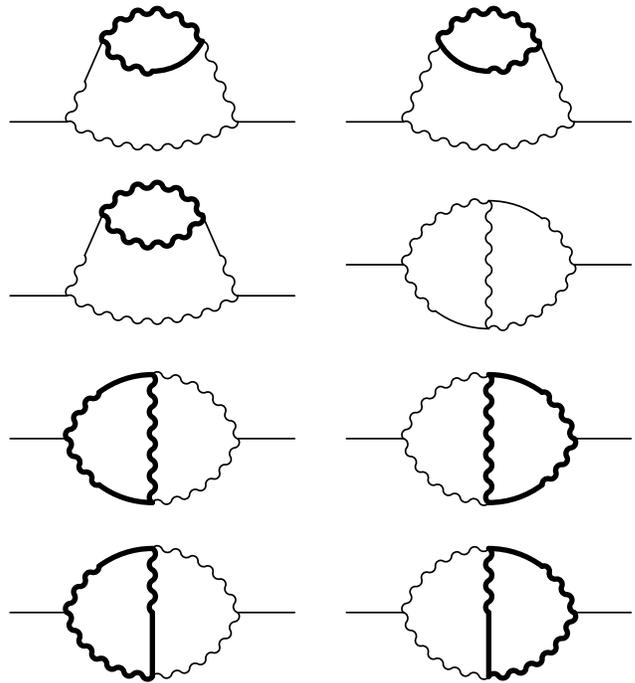}%
\caption{\label{2L1PI}Nonvanishing two-loop diagrams for the two-point 1PI
function $-\Gamma^{pp}$. Along with the overall divergence, each diagram has
at most one divergent subdiagram (bold lines).}
\end{figure}

Multiloop diagrams can be characterized by their overall UV divergence (as
all loop frequencies and wavenumbers go to infinity together) and their
subdivergences (as a subset of loop frequencies and wavenumbers go to
infinity while the rest remain finite). Overlapping divergences occur when
two or more divergent subdiagrams share a propagator; this is a definite
complication in evaluating a diagram, though not insurmountable \cite{Z96}.
Some two-loop diagrams for the 1PI function $-\Gamma^{\psi p}$ contain
overlapping divergences, so we choose to calculate the diagrams for
$-\Gamma^{pp}$, which fortunately do not (Fig.~\ref{2L1PI}). A three-point
1PI subdiagram is divergent only if its external lines are $\psi\psi p$, two
wiggly and one plain; Galilean invariance does not eliminate this
subdivergence when we treat each two-loop diagram separately, because the
finiteness of $-\Gamma^{\psi\psi p}_1$ results from a sum over distinct
one-loop diagrams. In Fig.~\ref{2L1PI} we have omitted one conceivable
diagram containing the 1PI subdiagram $-\Gamma^{\psi\psi}_1$, which vanishes
as noted in Sec.~\ref{Feynman}.

Previous two-loop calculations of similar complexity have been made for
different problems using diagrams of the same topology: the
Burgers--Kardar-Parisi-Zhang equation for interface growth \cite{FT94} and
the Navier--Stokes equation in more than two physical dimensions, where there
are fewer divergences \cite{AAKV03}.

\subsection{\label{Analytic}Analytic calculations}

We have programmed \textsc{mathematica} \cite{W99} to automate the steps in
evaluating each two-loop diagram. The loop frequencies are integrated one
after the other by adding the residues of all poles in the upper half-plane.
The resulting integrand, containing the external wavevector $\mathbf{q}$ and
the loop wavevectors $\mathbf{k}_{1,2}$, is expanded to $O(q^4)$ and averaged
over directions of $\mathbf{q}$ in the physical subspace. The numerator of
the integrand is now rife with the projector $\Theta_{ij}$, both from
averaging $q_i q_j$ and from applying Eq.~(\ref{epseps}) to cross products.
To eliminate $\Theta_{ij}$, we average the integrand over orientations of the
physical subspace within $d$-dimensional space: We introduce an orthonormal
physical basis, write $\Theta_{ij} = \hat a_i \hat a_j + \hat b_i \hat b_j$,
average over directions of $\hat{\mathbf{a}}$ in the $d - 1$ dimensions
perpendicular to $\hat{\mathbf{b}}$, and then average over directions of
$\hat{\mathbf{b}}$ in $d$-dimensional space. The result is a function of
$k_1^2$, $k_2^2$, and $\mathbf{k}_1 \cdot \mathbf{k}_2$ to be integrated over
$d^d k_1\, d^d k_2$.

Because we need only the divergent parts of the two-loop diagrams as
indicated in Eq.~(\ref{2Lpp}), we eliminate numerator terms that produce
neither an overall divergence nor a subdivergence by power counting. We seek
to integrate first over the loop wavevector (say $\mathbf{k}_1$) associated
with the subdivergence (if any). To simplify the denominator, Feynman
parameters \cite{Z96} are introduced, and $\mathbf{k}_1$ is translated by a
multiple of $\mathbf{k}_2$. After the numerator is isotropized, the
$\mathbf{k}_1$ integration is done analytically; there remains an integral
over Feynman parameters and over the magnitude $k_2$. From Eq.~(\ref{2Lpp}),
the integrand (excluding $g^3 q^4$) has dimension $-1 - 2\epsilon$. The part
that behaves like $k_2^{-1-2\epsilon}$ as $k_2 \to \infty$ is subtracted and
separately integrated over $k_2$, producing another $\epsilon^{-1}$ factor;
all the $\epsilon^{-2}$ terms arise here and are found analytically, but the
$\epsilon^{-1}$ terms involve intractable Feynman-parameter integrals.
Meanwhile, the remaining subtracted integral converges at $k_2 = \infty$ but
contains analytically integrable $\epsilon^{-1}$ poles from the
$\mathbf{k}_1$ integration.

Adding the divergent parts of the eight diagrams gives
\begin{align}
\Gamma^{pp}_2 = g^3 q^4 \biggl(&{-}\frac{1}{2048\pi^2 \epsilon^2}\nonumber\\
\label{an2Lpp}
&+ \frac{-2\ln(4\pi) + 2\gamma_{\text{E}} + 5 + X}{4096\pi^2 \epsilon} +
O(\epsilon^0)\biggr).
\end{align}
Here $X$ is a sum of Feynman-parameter integrals of complicated rational
functions with integer coefficients; thus we expect that $X$ may be a
rational number. Though we are unable to calculate $X$ analytically,
Eq.~(\ref{an2Lpp}) already displays some important features: If the
$\epsilon^{-2}$ term were different, the renormalization performed below
would become inconsistent \cite{T73}; and if the coefficients of
$\gamma_{\text{E}}$ and $\ln(4\pi)$ were different, these constants would
(contrary to expectation) appear in the two-loop $\beta$ function.

\subsection{\label{Numerical}Numerical calculations}

We have evaluated $X$ numerically by multidimensional Monte Carlo
integration, using the technique of importance sampling \cite{PTVF92} to
select more points in the ``corners'' of Feynman-parameter space (with one
parameter near $1$ and the others near $0$). This technique improves the
statistics because the integrands tend to diverge in these corners (but
slowly enough that the integrals converge). We treat the integrals that make
up $X$ separately, since they vary in complexity and in number of Feynman
parameters. To optimize the precision of the result for $X$ in a given
computation time $T$, we reason as follows. The contribution to $X$ from
diagram $i$, evaluated with $n_i$ sample points, is obtained with a precision
$s_i = \sigma_i n_i^{-1/2}$ in a time $t_i = \tau_i n_i$, where $\sigma_i$
and $\tau_i$ are characteristics of the integrand and the computer.
Constrained optimization shows that we achieve the best overall precision
\begin{equation}
S^2 = \sum_i s_i^2
\end{equation}
in the time
\begin{equation}
T = \sum_i t_i
\end{equation}
by choosing
\begin{equation}
n_i = \frac{T\sigma_i}{\tau_i^{1/2} \sum_j \sigma_j \tau_j^{1/2}}.
\end{equation}
Short runs are made to estimate $\sigma_i$ and $\tau_i$, and then the
high-precision integrations are performed with this plan. Our computations
for a total of $3.6 \times 10^8$ sample points yield
\begin{equation}
X = -3.995 \pm 0.005,
\end{equation}
strongly suggesting that the exact value is
\begin{equation}
X = -4.
\end{equation}

With the one-loop result (\ref{1Lpp}) and the two-loop result (\ref{an2Lpp}),
we have
\begin{multline}
\Gamma^{pp} = q^4 \biggl[g^{-1} + g \left(\frac{1}{32\pi\epsilon} -
\frac{\ln(g\rho/4\pi) + \gamma_{\text{E}}}{64\pi} + O(\epsilon)\right)\\
+ g^3 \left(-\frac{1}{2048\pi^2 \epsilon^2} + \frac{2\ln(g\rho/4\pi) +
2\gamma_{\text{E}} + 1}{4096\pi^2 \epsilon} + O(\epsilon^0)\right)\\
+ O(g^5)\biggr].
\end{multline}
By substituting Eq.~(\ref{2Lg}) and requiring a finite expression at
$\epsilon = 0$, we determine
\begin{equation}
\begin{aligned}
a_1 &= \frac{1}{32\pi}, & a_2 &= \frac{3}{2048\pi^2}, & a_2' &=
-\frac{1}{4096\pi^2}.
\end{aligned}
\end{equation}
The RG invariance of the bare coupling $g$ finally gives
\begin{equation}
\beta(\bar g) = -\tfrac{1}{2} \epsilon \bar g + \frac{\bar g^3}{32\pi} -
\frac{\bar g^5}{2048\pi^2} + O(\bar g^7).
\end{equation}
This is reminiscent of the $\beta$ function in four-dimensional $\phi^4$
theory, where similarly the one-loop term is positive and the two-loop term
is negative \cite{Z96}. A positive $\beta$ function that grows too quickly
(faster than linearly) at large coupling raises the question of whether the
coupling becomes infinite at a large but finite renormalization scale. In our
theory this would suggest an absolute limit on the extent of the scaling
range for any inverse cascade. To avoid such a fate, the expansion of the
$\beta$ function must contain many negative terms to slow its initial
superlinear growth. It is pleasing to find such a term already at two loops.

\section{\label{Cascade}Inverse-cascade range}

\subsection{\label{Phenom}Inverse-cascade phenomenology}

The initial prediction of the two-dimensional inverse energy cascade
\cite{K67} assumed zero friction and small but nonzero viscosity. Kinetic
energy, continually injected at high wavenumbers, is expected to cascade down
through a quasisteady inertial range extending to lower and lower wavenumbers
as time passes. Within this range, the assumption of scale invariance implies
the energy spectrum
\begin{equation}
\label{Ek}
E(k) = C\, \mathcal{E}^{2/3} k^{-5/3},
\end{equation}
where $C$ is the two-dimensional Kolmogorov constant and $\mathcal{E}$ is the
rate of energy injection per unit mass. Numerical simulations with zero
friction \cite{FS84,SY93,SY94} confirm this spectrum until the cascade
approaches the minimum wavenumber associated with a finite system. Both
nonstationary behavior and finite-size effects, however, lie outside our
theoretical framework, which treats a fluid of infinite size that has been
stirred for an infinite time. In the absence of friction, such a fluid has an
equipartition spectrum as shown in Sec.~\ref{FDT}.

When friction is introduced, it is natural to expect a mirror image of the
viscous energy dissipation at high wavenumbers in the three-dimensional
direct cascade: A stationary inverse cascade should develop, with dissipation
by friction at low wavenumbers $k \sim k_{\text{fr}}$ and with the spectrum
(\ref{Ek}) at wavenumbers $k \gg k_{\text{fr}}$ where dissipation is
unimportant. Dimensional analysis gives
\begin{equation}
k_{\text{fr}} = \mathcal{E}^{-1/2} \alpha^{3/2},
\end{equation}
where $\alpha$ is the friction coefficient \cite{BCV00,DG01}. Several
numerical simulations \cite{SY94,BCV00} and laboratory experiments
\cite{S86,PT98} confirm this picture. Other numerical studies obtain similar
results but are more difficult to relate to our framework, since they modify
the friction term by removing derivatives \cite{BDF95} or by applying
friction only below a cutoff wavenumber \cite{MV91}. We are less concerned
about the common practice of adding derivatives to the viscous term
(hyperviscosity), because such a modified term is irrelevant and the RG flow
should introduce a normal viscous term to replace it. But friction modified
with inverse derivatives (hypofriction) is certainly relevant and is believed
to alter the dynamics of the inverse cascade \cite{SY94,BCV00}. In an extreme
case, hypofriction with eight inverse Laplacians destroys the $k^{-5/3}$
spectrum \cite{B94}. Hypofriction is intended to confine dissipation
explicitly to low wavenumbers; ordinary linear friction accomplishes the same
thing more gently, but makes it difficult in practice to achieve an inertial
range \cite{DG01}. We have used linear friction because it is physically
realistic and leads to a local field theory.

For convenience in relating the observed inverse cascade to our field theory
with the action (\ref{Sfppsi}), we adopt units of time in which
\begin{equation}
g\nu \equiv D_0^{1/2} \nu^{-1/2} = 1,
\end{equation}
so that the rescalings in Eq.~(\ref{ppsit}) are trivial. Then the kinematic
viscosity is
\begin{equation}
\label{nug}
\nu = g^{-1},
\end{equation}
the friction coefficient is
\begin{equation}
\label{alpharho}
\alpha = \rho,
\end{equation}
and the force correlation is
\begin{equation}
\label{fcg}
D(k^2) = g^{-1} k^2.
\end{equation}

In contrast to the formal methods of dimensional regularization and minimal
subtraction, a simple UV cutoff renders our theory finite in a physically
meaningful way while still preserving the symmetries noted in
Sec.~\ref{Properties}. A cutoff would have been inconvenient for the two-loop
calculations of Sec.~\ref{TwoLoop}, but it is appropriate for understanding
experimental and numerical results on the inverse cascade. The local,
renormalizable action (\ref{Sfppsi}) applies only with a UV cutoff $\Lambda$
below the wavenumbers of the external forcing \cite{FNS77}. A corresponding
renormalization prescription is obtained by staying in exactly two spatial
dimensions and writing the coupling $g$ in Eq.~(\ref{Sfppsi}) as a
cutoff-dependent quantity $\hat g(\Lambda)$, such that the long-distance
behavior is independent of $\Lambda$. No cutoff dependence is needed for
$\rho$, because the friction term is not renormalized, as we saw in
Sec.~\ref{OneLoop}.

Special properties of minimal subtraction allowed us to conclude
(Sec.~\ref{FDT}) that in that scheme the anomalous dimensions vanish and
$\beta(\bar g)$ depends only on $\bar g$. We may expect that in the cutoff
scheme these statements remain approximately true, at least for small values
of the dimensionless friction parameter $\Lambda^{-2} \rho$. Because the
$\beta$ function is scheme independent through two loops \cite{Z96}, we have
\begin{equation}
\label{Lbeta}
\Lambda \frac{d\hat g}{d\Lambda} \equiv \beta(\hat g) = \frac{\hat
g^3}{32\pi} - \frac{\hat g^5}{2048\pi^2} + O(\hat g^7).
\end{equation}
This result is directly useful for weak coupling, but can only hint at the
possible strong-coupling behavior. We will now show that a specific
strong-coupling form of the $\beta$ function is required for a fully
developed inverse cascade, and that it can be naturally interpolated with the
two-loop result (\ref{Lbeta}).

\subsection{\label{Strong}Strong-coupling behavior}

The key condition for the ideal inverse cascade is that the dissipation of
energy is dominated by friction and is almost totally confined to low
wavenumbers. This means that the dissipation rate
\begin{equation}
\label{diss}
\mathcal{E} = 2\rho \int_0^\Lambda dk\, E(k) = 2\rho \int_0^\Lambda dk\,
\frac{\pi k}{(2\pi)^2} k^2 \langle\psi\psi\rangle_=
\end{equation}
should be independent of the UV cutoff $\Lambda$, because a change in
$\Lambda$ produces neither a renormalization of $\rho$ nor a rescaling of
$\psi$. Under stationary conditions, $\mathcal{E}$ equals the rate of energy
injection, given by \cite{AAV99}
\begin{equation}
\label{EgL}
\mathcal{E} = \int_0^\Lambda dk\, \frac{\pi k}{(2\pi)^2}\, D(k^2) =
\frac{\Lambda^4}{16\pi\, \hat g(\Lambda)},
\end{equation}
in terms of the force correlation (\ref{fcg}). For $\mathcal{E}$ to be
independent of $\Lambda$, we must have the strong-coupling behavior
\begin{equation}
\hat g(\Lambda) \simeq \text{const} \times \Lambda^4,
\end{equation}
corresponding to the asymptotically linear $\beta$ function
\begin{equation}
\begin{aligned}
\beta(\hat g) &\simeq 4\hat g & (\hat g &\to \infty).
\end{aligned}
\end{equation}
We now attempt to connect this form with the two-loop $\beta$ function
(\ref{Lbeta}).

Perturbative expansions such as Eq.~(\ref{Lbeta}) are usually divergent, but
a Borel transformation is expected to produce a finite radius of convergence
about zero coupling \cite{Z96}. We see that the true expansion parameter is
$\hat g^2$, and the alternatively normalized action (\ref{Sippsi}) makes it
clear that the theory is unstable for $\hat g^2 < 0$. Thus we write
\begin{equation}
\label{BT}
\frac{\beta(\hat g)}{\hat g} \equiv A(\hat g^2) = \int_0^\infty
\frac{dz}{\hat g^2}\, B(z) \exp\frac{-z}{\hat g^2}.
\end{equation}
For the perturbation series
\begin{equation}
A(\hat g^2) = \sum_{n=1}^\infty A_n \hat g^{2n},
\end{equation}
the Borel transform is
\begin{equation}
B(z) = \sum_{n=1}^\infty \frac{A_n}{n!} z^n.
\end{equation}
In our case,
\begin{equation}
\label{Bser}
B(z) = \frac{z}{32\pi} - \frac{z^2}{4096\pi^2} + O(z^3).
\end{equation}

The Borel transform often has poles on the negative real axis associated with
instantons \cite{Z96}, but $\beta(\hat g)$ is well-defined from
Eq.~(\ref{BT}) as long as $B(z)$ is regular on the positive real axis. A
simple and suitable rational (Pad\'e) approximant to $B(z)$ is
\begin{equation}
\begin{aligned}
B(z) &\simeq \frac{z}{32\pi + yz} & (y &\ge 0);
\end{aligned}
\end{equation}
fortunately, the choice $y = \frac{1}{4}$ agrees with Eq.~(\ref{Bser}). We
thus take
\begin{equation}
B(z) \simeq \frac{4z}{128\pi + z},
\end{equation}
but we do not here attempt to study the possible instanton solutions
corresponding to the pole at $z = -128\pi$. Eq.~(\ref{BT}) then gives
precisely the desired asymptotic behavior
\begin{equation}
\begin{aligned}
\beta(\hat g) &\simeq 4\hat g & (\hat g &\to \infty).
\end{aligned}
\end{equation}

We now ask whether the observed $k^{-5/3}$ energy spectrum (\ref{Ek}) is
consistent with our theory, although we are unable to derive it
systematically. We conjecture that strong-coupling effects produce the
spectrum
\begin{equation}
\label{Ek1}
E(k) \sim \mathcal{E}^{2/3} k^{-5/3}
\end{equation}
for
\begin{equation}
k \gtrsim k_{\text{fr}} = \mathcal{E}^{-1/2} \rho^{3/2}.
\end{equation}
From Eq.~(\ref{EgL}), the running coupling is
\begin{equation}
\hat g(\Lambda) \sim \mathcal{E}^{-1} \Lambda^4.
\end{equation}
For nonperturbative effects to be operative down to the wavenumber
$k_{\text{fr}}$, we must have
\begin{equation}
\hat g(k_{\text{fr}}) \sim (\mathcal{E}^{-1} \rho^2)^3 \gtrsim 1.
\end{equation}
In the borderline case where $\hat g(k_{\text{fr}}) \sim 1$, we have
$k_{\text{fr}} \sim \mathcal{E}^{1/4} \sim \rho^{1/2}$, and we can match
Eq.~(\ref{Ek1}) in order of magnitude with the perturbative energy spectrum
(\ref{Ekp}):
\begin{equation}
E(k_{\text{fr}}) \sim \mathcal{E}^{2/3} k_{\text{fr}}^{-5/3} \sim
\mathcal{E}^{1/4} \sim \frac{k_{\text{fr}}^3}{\rho + k_{\text{fr}}^2}.
\end{equation}
Even when $\hat g(k_{\text{fr}}) \gg 1$, we may guess from Eq.~(\ref{Ekp})
that in order of magnitude
\begin{equation}
E(k_{\text{fr}}) \sim \frac{k_{\text{fr}}^3}{\rho\, \hat g(k_{\text{fr}}) +
k_{\text{fr}}^2} \sim \mathcal{E}^{3/2} \rho^{-5/2},
\end{equation}
and this again matches Eq.~(\ref{Ek1}) at $k_{\text{fr}}$.

In the case $\hat g(k_{\text{fr}}) \gg 1$, the running coupling eventually
becomes $\sim 1$ at a lower wavenumber
\begin{equation}
k_g \sim \mathcal{E}^{1/4},
\end{equation}
below which perturbation theory is applicable and the energy spectrum is
given roughly by Eq.~(\ref{Ekp1}). Thus we envision a varied but continuous
behavior of the energy spectrum in the different regions:
\begin{equation}
E(k) \sim
\begin{cases}
\rho^{-1} k^3 & (k \lesssim \mathcal{E}^{1/4}),\\
\mathcal{E} \rho^{-1} k^{-1} & (\mathcal{E}^{1/4} \lesssim k \lesssim
\mathcal{E}^{-1/2} \rho^{3/2}),\\
\mathcal{E}^{2/3} k^{-5/3} & (k \gtrsim \mathcal{E}^{-1/2} \rho^{3/2}).
\end{cases}
\end{equation}
As a further check, we note that the contribution to the energy dissipation
rate (\ref{diss}) for each of the three regions is $\sim \mathcal{E}$ (up to
a logarithmic factor for the middle region). This suggests that a fully
developed stationary inverse cascade has several distinct dissipation ranges.
Testing by experiments and simulations is not straightforward, because
finite-size effects may become important before the lower-wavenumber ranges
are reached.

\subsection{\label{Flux}Energy flux and third moment}

We have described the ideal inverse cascade in terms of external forcing
confined to high wavenumbers (assumed in Sec.~\ref{Relevance}) and energy
dissipation practically confined to low wavenumbers (made plausible in Secs.\
\ref{RG} and \ref{Strong}). This separation is equivalent to the conventional
criterion of an inertial range with a constant energy flux. However, unlike
the energy spectrum (a simple correlation function), the energy flux in
wavenumber space is not invariant under the RG flow and is not given
straightforwardly by our local field theory. Renormalization introduces the
forcing (\ref{fcg}), which is nonzero even for wavenumbers in the inertial
range and so makes the flux appear nonconstant. This effective forcing simply
represents the energy transfer from wavenumbers $k > \Lambda$ that have been
integrated out.

We can resolve the flux problem by working instead in physical space, where
the effective forcing is a differentiated delta function that is zero at
finite distances, including the inertial range of lengths. Thus we inquire
whether the physical-space energy flux \cite{F95}
\begin{equation}
\label{er}
\varepsilon(r) = -\tfrac{1}{4} \bm{\nabla}_{\mathbf{r}} \cdot \langle
|\delta\mathbf{v}(\mathbf{r})|^2 \delta\mathbf{v}(\mathbf{r}) \rangle
\end{equation}
is preserved under renormalization. The flux is given in terms of the third
moment of the velocity increment
\begin{equation}
\delta\mathbf{v}(\mathbf{r}) = \mathbf{v}(\mathbf{x}+\mathbf{r}) -
\mathbf{v}(\mathbf{x})
\end{equation}
and is independent of $\mathbf{x}$ by homogeneity. Under stationary
conditions, the Navier--Stokes equation (\ref{NSv}) yields \cite{B99}
\begin{align}
\varepsilon(r) &= (\nu\nabla_{\mathbf{r}}^2 - \alpha) \langle
\mathbf{v}(\mathbf{x}) \cdot \mathbf{v}(\mathbf{x}+\mathbf{r}) \rangle +
\tfrac{1}{2} \hat C(r)\nonumber\\
\label{er1}
&= [\hat g(\Lambda)^{-1} \nabla_{\mathbf{r}}^2 - \rho] \langle
\mathbf{v}(\mathbf{x}) \cdot \mathbf{v}(\mathbf{x}+\mathbf{r}) \rangle,
\end{align}
where $\hat C(r)$ is the vanishing physical-space force correlation, and we
have used Eqs.\ (\ref{nug}) and (\ref{alpharho}).

The right-hand side of Eq.~(\ref{er1}) can be interpreted as minus the rate
of energy dissipation at length scales larger than $r$. Since $\langle
\mathbf{v}(\mathbf{x}) \cdot \mathbf{v}(\mathbf{x}+\mathbf{r}) \rangle$ is an
RG-invariant correlation function, only the viscosity $\hat g(\Lambda)^{-1}$
introduces cutoff dependence. In accordance with our argument in
Sec.~\ref{Strong}, as long as the energy dissipation is dominated by
friction, the dissipation rate is independent of $\Lambda$, giving a
well-defined energy flux
\begin{equation}
\varepsilon(r) = -\rho \langle \mathbf{v}(\mathbf{x}) \cdot
\mathbf{v}(\mathbf{x}+\mathbf{r}) \rangle
\end{equation}
proportional to the Fourier transform of the energy spectrum. And if this
spectrum is almost entirely concentrated at low wavenumbers $k \lesssim
k_{\text{fr}}$, it follows that in the inertial range ($r \ll
k_{\text{fr}}^{-1}$) the flux is constant,
\begin{equation}
\varepsilon(r) \simeq \varepsilon(0) = -\rho \langle v^2 \rangle =
-\mathcal{E}.
\end{equation}

\subsection{\label{Intermitt}Intermittency}

A striking feature observed in the inverse cascade is the approximate scale
invariance of inertial-range velocity correlations (structure functions). The
$k^{-5/3}$ energy spectrum (\ref{Ek}) gives for the second moment \cite{F95}
\begin{equation}
\label{vv}
\langle |\delta\mathbf{v}(\mathbf{r})|^2 \rangle \propto \mathcal{E}^{2/3}
r^{2/3},
\end{equation}
and the constant energy flux (\ref{er}) gives for the third moment \cite{B99}
\begin{equation}
\label{vvv}
\langle |\delta\mathbf{v}(\mathbf{r})|^2 \delta\mathbf{v}(\mathbf{r}) \rangle
\propto \mathcal{E} \mathbf{r}.
\end{equation}
In our field theory, since the anomalous dimensions vanish, such moments can
be written in the form
\begin{equation}
\label{vn}
\langle (\delta v)^n \rangle = r^{-n} f_n\bm{(}\hat g(r^{-1}), \rho
r^2\bm{)},
\end{equation}
where $r^{-n}$ is the kinematic scaling and $f_n$ is a function of the
dimensionless running couplings. Eq.~(\ref{vn}) should reduce to the observed
forms in the inertial range
\begin{equation}
\label{inr}
r \ll k_{\text{fr}}^{-1} = \mathcal{E}^{1/2} \rho^{-3/2},
\end{equation}
subject to the condition for a fully developed cascade,
\begin{equation}
\label{gkfr}
\hat g(k_{\text{fr}}) \sim (\mathcal{E}^{-1} \rho^2)^3 \gg 1.
\end{equation}
Conditions (\ref{inr}) and (\ref{gkfr}) are equivalent to
\begin{equation}
\label{cond}
(\rho r^2)^{-2} \ll \hat g(r^{-1}) \ll (\rho r^2)^{-3}.
\end{equation}

To achieve complete scale invariance of inertial-range structure functions
(absence of intermittency), we must have
\begin{equation}
\label{fn}
f_n\bm{(}\hat g(r^{-1}), \rho r^2\bm{)} \propto \hat g(r^{-1})^{-n/3}
\end{equation}
for the range of arguments (\ref{cond}); then $f_n \propto \mathcal{E}^{n/3}
r^{4n/3}$, and
\begin{equation}
\langle (\delta v)^n \rangle \propto \mathcal{E}^{n/3} r^{n/3}.
\end{equation}
Eq.~(\ref{fn}) requires a specific nonperturbative behavior of velocity
correlations at strong coupling, mimicking the effect of an anomalous
dimension. RG theory alone does not constrain the functions $f_n$. The
well-established second moment (\ref{vv}) and third moment (\ref{vvv})
strongly suggest that an exact calculation in our theory would yield the
required behavior of $f_2$ and $f_3$. But in the absence of a field-theoretic
reason why Eq.~(\ref{fn}) should persist for $n \ge 4$, high-order structure
functions may generically be expected to violate scale invariance and produce
intermittency. In sum, we would not be at all surprised by observations of
intermittency in the inverse cascade, but a seeming total absence of
intermittency would raise questions about unknown properties of our theory
that enforce effective scale invariance.

Some numerical studies of higher moments \cite{SY93,SY94} deal with a
nonstationary inverse cascade and find no signs of intermittency. Such
results are not directly relevant to this paper but have been explained
theoretically \cite{Y99} based on the growth of the inertial range with time.
For the stationary inverse cascade, laboratory experiments \cite{PT98,DBPT01}
reveal no evidence of intermittency. The results of stationary numerical
simulations, however, are mixed: A study using linear friction \cite{BCV00}
confirms the absence of intermittency, but a study using hypofriction with
one inverse Laplacian \cite{BDF95} obtains intermittency that is described as
strong and as similar to that observed in the three-dimensional direct
cascade. Surprisingly, results from the latter simulation are also presented
in a subsequent paper \cite{DBPT01} where the intermittency is described as
insignificant. The evidence on intermittency in the stationary inverse
cascade is unclear, and further numerical studies would be useful to resolve
the question. Results exhibiting strong intermittency are most natural, from
the viewpoint of our field theory.

The mild hypofriction used in the numerical study obtaining strong
intermittency \cite{BDF95} is unlikely to alter the qualitative structure of
our theory. While the inverse Laplacian renders the initial Navier--Stokes
equation (\ref{NSv}) nonlocal, the differentiated form (\ref{NSpsi}) and the
field-theory action (\ref{Sfppsi}) are still local in terms of the stream
function $\psi$. The hypofriction term certainly violates Galilean
invariance, but this symmetry holds for zero hypofriction and so its
implications for the RG functions persist in minimal subtraction even when
hypofriction is added. The same arguments used for linear friction in this
paper suggest that intermittency should be expected with hypofriction as
well. It would be interesting to perform a direct numerical study of the
effect of modified friction on intermittency in the stationary inverse
cascade.

\section{\label{Burgers}Generalized Burgers equation}

\subsection{\label{Continuation}Dimensional continuation}

As an example of an alternative statistical fluid model to which our RG
methods can be applied but which exhibits different behavior, let us briefly
consider the one-dimensional Burgers equation \cite{YS88}
\begin{equation}
\label{B1}
\dot v + v\nabla v - \nu\nabla^2 v + \alpha v = f.
\end{equation}
Here $v$ is the velocity field of a fluid without pressure, $f$ is the force
per unit mass, $\nu$ is the kinematic viscosity, and $\alpha$ is the friction
coefficient (not normally included in the Burgers equation but useful in
controlling the long-distance behavior). As with the two-dimensional
incompressible fluid, we assume that the forcing is confined to a band of
high wavenumbers, and we study the response at lower wavenumbers.

FNS \cite{FNS77} found that the UV-stirred Burgers equation, like the
Navier--Stokes equation, has a dimensionless coupling in two spatial
dimensions and can be analyzed by an $\epsilon$-expansion in $d = 2 -
\epsilon$. But their continuation of the Burgers equation to $d > 1$ does not
preserve important one-dimensional properties, and their $\epsilon$-expansion
is not fully consistent \cite{FNS77,YS88}. With $\nu = \alpha = f = 0$,
Eq.~(\ref{B1}) yields conservation of the ``energy''
\begin{equation}
E = \tfrac{1}{2} \int dx\, v^2.
\end{equation}
This $E$ is not proportional to the physical energy of a pressure-free fluid,
because it does not account for the varying density; nevertheless, $E$ is
conserved in $d = 1$, and leads to a fluctuation-dissipation theorem (FDT).
It is not easy, however, to generalize Eq.~(\ref{B1}) to $d > 1$ so that a
similar ``energy'' is conserved, while maintaining other key properties such
as Galilean invariance. This is the task we now address.

If we neglect dissipation and forcing, the conservation of $E$ in $d = 1$
follows from the relation
\begin{equation}
0 = v\dot v + \tfrac{1}{3} \nabla(v^3) \equiv v(\dot v + v\nabla v).
\end{equation}
The analogue in $d > 1$ that would yield conservation of
\begin{equation}
E = \tfrac{1}{2} \int d^d x\, |\mathbf{v}|^2
\end{equation}
is
\begin{align}
0 &= \mathbf{v} \cdot \dot{\mathbf{v}} + A \bm{\nabla} \cdot (v^2
\mathbf{v})\nonumber\\
&\equiv \mathbf{v} \cdot [\dot{\mathbf{v}} + A (\bm{\nabla} \cdot \mathbf{v})
\mathbf{v} + (A - B) \bm{\nabla}(v^2) + 2B (\mathbf{v} \cdot \bm{\nabla})
\mathbf{v}]\nonumber\\
&\equiv \mathbf{v} \cdot \mathbf{w}.
\end{align}
Unfortunately, whatever the choice of the constants $A$ and $B$, the quantity
$\mathbf{w}$ is not Galilean covariant and is not a suitable generalization
of $(\dot v + v\nabla v)$.

To remedy this problem, we impose the potential-flow condition
\begin{equation}
\mathbf{v} = \bm{\nabla}\psi,
\end{equation}
as is commonplace when considering a multidimensional Burgers equation. Then,
because
\begin{equation}
\dot E = \int d^d x\, \mathbf{v} \cdot \mathbf{w} = \int d^d x\, \psi
(-\bm{\nabla} \cdot \mathbf{w}),
\end{equation}
a scalar equation of motion can conserve
\begin{equation}
E = \tfrac{1}{2} \int d^d x\, \nabla_i \psi \nabla_i \psi.
\end{equation}
We take
\begin{align}
0 = -\bm{\nabla} \cdot \mathbf{w} \equiv {}&{-}\nabla^2 \dot\psi - A\nabla^2
\psi \nabla^2 \psi - 2A\nabla_i \nabla_j \psi \nabla_i \nabla_j
\psi\nonumber\\
&- 3A\nabla_i \psi \nabla_i \nabla^2 \psi.
\end{align}
If $A = \tfrac{1}{3}$, this expression is covariant under the Galilean
transformation
\begin{equation}
\psi(t, \mathbf{x}) \to \psi(t, \mathbf{x} + \mathbf{u}t) - \mathbf{u} \cdot
\mathbf{x},
\end{equation}
which induces
\begin{equation}
\mathbf{v}(t, \mathbf{x}) \to \mathbf{v}(t, \mathbf{x} + \mathbf{u}t) -
\mathbf{u}.
\end{equation}

Upon restoring dissipation and forcing, we therefore propose
\begin{align}
&{-}\nabla^2 \dot\psi - \nabla_i \psi \nabla_i \nabla^2 \psi - \tfrac{1}{3}
\nabla^2 \psi \nabla^2 \psi\nonumber\\
\label{Bd}
&- \tfrac{2}{3} \nabla_i \nabla_j \psi \nabla_i \nabla_j \psi + \nu\nabla^4
\psi - \alpha\nabla^2 \psi = \eta
\end{align}
as a fully satisfactory multidimensional Burgers equation that reduces to
Eq.~(\ref{B1}) in $d = 1$. Eq.~(\ref{Bd}) is formally very similar to the
incompressible stream-function equation (\ref{NSpsi}), and we will follow our
previous analysis closely.

\subsection{\label{OneLoopB}One-loop renormalization}

As in Sec.~\ref{Framework}, we assume Gaussian forcing and introduce a path
integral with the action
\begin{alignat}{2}
S = {}&{\int dt}\, d^d x\, \bigl[\tfrac{1}{2} (-\nabla^2 p)\,
D(-\nabla^2)p\span\span\nonumber\\
&+ ip(&&{-}\nabla^2 \dot\psi - \nabla_i \psi \nabla_i \nabla^2 \psi -
\tfrac{1}{3} \nabla^2 \psi \nabla^2 \psi\nonumber\\
&&&- \tfrac{2}{3} \nabla_i \nabla_j \psi \nabla_i \nabla_j \psi + \nu\nabla^4
\psi - \alpha\nabla^2 \psi)\bigr]\nonumber\\
= {}&{\int dt}\, d^d x\, \bigl[\tfrac{1}{2} (-\nabla^2 p)\, D(-\nabla^2)p +
i\nu p\nabla^4 \psi - i\alpha p\nabla^2 \psi\span\span\nonumber\\
&- ip\nabla^2 \dot\psi - i\nabla_i \nabla_j p\, \bigl(\tfrac{1}{6}
\delta_{ij} \nabla_k \psi \nabla_k \psi + \tfrac{1}{3} \nabla_i \psi \nabla_j
\psi\bigr)\bigr],\span\span
\end{alignat}
where we have again integrated the $p\psi\psi$ term by parts. Analysis of
dimensions in $d = 2 - \epsilon$ proceeds as before, since all terms contain
the same numbers of derivatives as for the incompressible fluid. The final
Burgers action, to be compared with Eq.~(\ref{Sfppsi}), is
\begin{align}
S = {}&{\int dt}\, d^d x\, \bigl[\tfrac{1}{2} g^{-1} \nabla^2 p \nabla^2 p +
ig^{-1} p\nabla^4 \psi - i\rho p\nabla^2 \psi\nonumber\\
&- ip\nabla^2 \dot\psi - i\nabla_i \nabla_j p\, \bigl(\tfrac{1}{6}
\delta_{ij} \nabla_k \psi \nabla_k \psi + \tfrac{1}{3} \nabla_i \psi \nabla_j
\psi\bigr)\bigr].
\end{align}
The scaling dimensions are again given by Eq.~(\ref{deps}).

The only change to the Feynman rules in Fig.~\ref{FeynRules} is the vertex
factor, which now becomes
\begin{equation}
-\Gamma^{\psi\psi p}_0 = \tfrac{1}{3} i |\mathbf{k}_1 + \mathbf{k}_2|^2
\mathbf{k}_1 \cdot \mathbf{k}_2 + \tfrac{2}{3} i (k_1^2 + \mathbf{k}_1 \cdot
\mathbf{k}_2) (k_2^2 + \mathbf{k}_1 \cdot \mathbf{k}_2).
\end{equation}
Thanks to conservation of the ``energy'' $E$, the FDT is obtained just as in
Sec.~\ref{FDT}, and together with Galilean invariance it implies that the
anomalous dimensions vanish. Thus, to find the new one-loop $\beta$ function,
we need only recompute one of the diagrams in Fig.~\ref{1L1PI}. We calculate
\begin{equation}
\label{1LpsipB}
\Gamma^{\psi p}_1 = igq^4 \left(\frac{1}{32\pi\epsilon} -
\frac{\ln(g\rho/4\pi) + \gamma_{\text{E}} + \frac{8}{9}}{64\pi} +
O(\epsilon)\right),
\end{equation}
to be compared with Eq.~(\ref{1Lpsip}). Curiously, the $O(\epsilon^{-1})$
term is exactly the same and again leads to the one-loop $\beta$ function
\begin{equation}
\label{betaB}
\beta(\bar g) = -\tfrac{1}{2} \epsilon \bar g + \frac{\bar g^3}{32\pi} +
O(\bar g^5).
\end{equation}
But the $O(\epsilon^0)$ term in Eq.~(\ref{1LpsipB}), which would enter a
calculation of the two-loop $\beta$ function, differs from
Eq.~(\ref{1Lpsip}), and we have no reason to expect an identity between the
$\beta$ functions beyond one loop.

Our multidimensional Burgers equation (\ref{Bd}) appears not to have been
investigated previously, and it is possible that numerical simulations in two
or three dimensions could reveal interesting and unexpected behavior. The
focus here, however, is on the extrapolation to one dimension ($\epsilon =
1$), where we seek to explain the known behavior \cite{YS88} of the ordinary
Burgers equation (\ref{B1}). If $\epsilon$ is positive and small, then the
$\beta$ function (\ref{betaB}) has a nontrivial IR-stable fixed point at
\begin{equation}
\bar g_*^2 = 16\pi\epsilon + O(\epsilon^2).
\end{equation}
By contrast, when a simpler continuation of the Burgers equation was used,
the fixed point appeared to be IR-unstable near two dimensions
\cite{FNS77,YS88}. From our results it is natural to expect an IR-stable
strong-coupling fixed point in one dimension ($\epsilon = 1$).

\subsection{\label{Comparison}Comparison with inverse-cascade model}

Due to the nontrivial fixed point, the response of the one-dimensional
Burgers equation to UV forcing is different from that of the two-dimensional
Navier--Stokes equation---even in the absence of friction. At wavenumbers low
enough that the fixed point is reached, Burgers correlation functions (at
equal or unequal times) exhibit purely kinematic scaling, since all anomalous
dimensions vanish. For example, from Eq.~(\ref{deps}), the scaling of time
and frequency is given by
\begin{equation}
\begin{aligned}
d_t &= -2 + \tfrac{1}{2} \epsilon = -\tfrac{3}{2}, & \omega &\propto k^{3/2}.
\end{aligned}
\end{equation}
Meanwhile, in both models, the FDT implies that the energy spectrum (an
equal-time correlation function) obeys equipartition. These conclusions agree
with previous analytic \cite{FNS77} and numerical \cite{YS88} studies of the
frictionless one-dimensional Burgers equation.

We have argued that the inclusion of friction dramatically alters the
behavior of the two-dimensional Navier--Stokes model, from equipartition to a
stationary inverse cascade. This is plausible only in the presence of a
large, rapidly flowing coupling $\hat g$. We noted in Sec.~\ref{Cascade} how
this RG flow could give rise to a constant energy flux and a $k^{-5/3}$
spectrum. No such profound effect is expected for friction in the
one-dimensional Burgers equation. Correlation functions will be modified at
very low wavenumbers $\lesssim \rho^{2/3}$, but as long as $\hat g$ remains
close to the fixed point, higher wavenumbers will retain the equipartition
spectrum.

An open question concerns the strong-coupling behavior of our generalized
Burgers equation (\ref{Bd}) in two dimensions with friction. If the $\beta$
function happens to grow asymptotically in the same way as the Navier--Stokes
one, then the arguments of Sec.~\ref{Cascade} can be repeated and it is at
least possible that this Burgers model could exhibit an inverse energy
cascade. If true, this should also be evident if the equation is studied
numerically. It is unclear, however, whether this multidimensional
Galilean-invariant scalar field theory has a direct physical application.

\section{\label{Discussion}Discussion}

We have presented a statistical field theory capable of describing the
stationary inverse energy cascade in two-dimensional incompressible
turbulence, and computed the RG functions through two loops. By contrast with
previous RG studies of turbulence, we have taken advantage of the nature of
the inverse cascade, with external forcing confined to high wavenumbers, to
work with a conventional local field theory. The consequences of Galilean
invariance and the fluctuation-dissipation theorem have been systematically
derived based on the underlying symmetries of the action. After taking these
symmetries into account, we have evaluated the necessary two-loop diagrams in
dimensional regularization to obtain the two-loop $\beta$ function
\begin{equation}
\beta(\bar g) = \frac{\bar g^3}{32\pi} - \frac{\bar g^5}{2048\pi^2} + O(\bar
g^7),
\end{equation}
which is independent of the renormalization scheme to precisely this order.

Because the anomalous dimensions vanish identically in minimal subtraction,
no RG fixed point can yield the observed $k^{-5/3}$ energy spectrum of the
inverse cascade. Instead, we have found that the inverse cascade could
plausibly arise from nonperturbative strong-coupling effects in the presence
of friction. The apparent anomalous dimension of the velocity must arise from
the rapid RG flow of the coupling. Cutoff independence of the energy
dissipation rate (or constancy of the energy flux) requires the
strong-coupling behavior
\begin{equation}
\begin{aligned}
\beta(\hat g) &\simeq 4\hat g & (\hat g &\to \infty),
\end{aligned}
\end{equation}
and we have also obtained this in a heuristic way from a Borel transformation
of the two-loop $\beta$ function. The observed energy spectrum in the
inertial range has been matched with perturbative results at the wavenumbers
where dissipation becomes important. Inertial-range intermittency (violation
of scale invariance) is generically expected because the coupling flows
rapidly with scale, but the evidence on intermittency from numerical
simulations is mixed. On the other hand, a similar RG analysis of the
one-dimensional Burgers equation confirms the simpler behavior of that model,
controlled by a strong-coupling fixed point.

Our greatest difficulty is that the inverse cascade appears to be
intrinsically a nonperturbative phenomenon. To make quantitative predictions,
such as the exponent of the energy spectrum or the value of the Kolmogorov
constant, it may be useful to combine a high-order perturbative calculation
with an appropriate resummation method, as in our analysis of the
strong-coupling $\beta$ function. Of course evaluating additional diagrams
with two or more loops will be very challenging, especially if the finite
parts are needed. At the present stage, the most intriguing prediction of our
theory is that substantial intermittency should be expected in the stationary
inverse cascade. It would be helpful to have more robust and consistent
results from experiments and simulations to determine whether this
expectation is realized, and if it is not, to identify the field-theoretic
explanation.

\begin{acknowledgments}

I thank A. M. Polyakov, S. L. Sondhi, H. L. Verlinde, and V. Yakhot for
helpful discussions.

This material is based upon work supported by the National Science Foundation
under Grant No.\ 0243680. Any opinions, findings, and conclusions or
recommendations expressed in this material are those of the author and do not
necessarily reflect the views of the National Science Foundation.

\end{acknowledgments}


\begin{thebibliography}{34}
\expandafter\ifx\csname natexlab\endcsname\relax\def\natexlab#1{#1}\fi
\expandafter\ifx\csname bibnamefont\endcsname\relax
  \def\bibnamefont#1{#1}\fi
\expandafter\ifx\csname bibfnamefont\endcsname\relax
  \def\bibfnamefont#1{#1}\fi
\expandafter\ifx\csname citenamefont\endcsname\relax
  \def\citenamefont#1{#1}\fi
\expandafter\ifx\csname url\endcsname\relax
  \def\url#1{\texttt{#1}}\fi
\expandafter\ifx\csname urlprefix\endcsname\relax\def\urlprefix{URL }\fi
\providecommand{\bibinfo}[2]{#2}
\providecommand{\eprint}[2][]{\url{#2}}

\bibitem[{\citenamefont{Kraichnan}(1967)}]{K67}
\bibinfo{author}{\bibfnamefont{R.~H.} \bibnamefont{Kraichnan}},
  \bibinfo{journal}{Phys. Fluids} \textbf{\bibinfo{volume}{10}},
  \bibinfo{pages}{1417} (\bibinfo{year}{1967}).

\bibitem[{\citenamefont{Zinn-Justin}(1996)}]{Z96}
\bibinfo{author}{\bibfnamefont{J.}~\bibnamefont{Zinn-Justin}},
  \emph{\bibinfo{title}{Quantum Field Theory and Critical Phenomena}}
  (\bibinfo{publisher}{Oxford University Press}, \bibinfo{address}{Oxford},
  \bibinfo{year}{1996}), \bibinfo{edition}{3rd} ed.

\bibitem[{\citenamefont{Frisch}(1995)}]{F95}
\bibinfo{author}{\bibfnamefont{U.}~\bibnamefont{Frisch}},
  \emph{\bibinfo{title}{Turbulence: The Legacy of A. N. Kolmogorov}}
  (\bibinfo{publisher}{Cambridge University Press},
  \bibinfo{address}{Cambridge}, \bibinfo{year}{1995}).

\bibitem[{\citenamefont{Smith and Yakhot}(1993)}]{SY93}
\bibinfo{author}{\bibfnamefont{L.~M.} \bibnamefont{Smith}} \bibnamefont{and}
  \bibinfo{author}{\bibfnamefont{V.}~\bibnamefont{Yakhot}},
  \bibinfo{journal}{Phys. Rev. Lett.} \textbf{\bibinfo{volume}{71}},
  \bibinfo{pages}{352} (\bibinfo{year}{1993}).

\bibitem[{\citenamefont{Smith and Yakhot}(1994)}]{SY94}
\bibinfo{author}{\bibfnamefont{L.~M.} \bibnamefont{Smith}} \bibnamefont{and}
  \bibinfo{author}{\bibfnamefont{V.}~\bibnamefont{Yakhot}},
  \bibinfo{journal}{J. Fluid Mech.} \textbf{\bibinfo{volume}{274}},
  \bibinfo{pages}{115} (\bibinfo{year}{1994}).

\bibitem[{\citenamefont{Yakhot}(1999)}]{Y99}
\bibinfo{author}{\bibfnamefont{V.}~\bibnamefont{Yakhot}},
  \bibinfo{journal}{Phys. Rev. E} \textbf{\bibinfo{volume}{60}},
  \bibinfo{pages}{5544} (\bibinfo{year}{1999}).

\bibitem[{\citenamefont{Babiano et~al.}(1995)\citenamefont{Babiano, Dubrulle,
  and Frick}}]{BDF95}
\bibinfo{author}{\bibfnamefont{A.}~\bibnamefont{Babiano}},
  \bibinfo{author}{\bibfnamefont{B.}~\bibnamefont{Dubrulle}},
\bibnamefont{and}
  \bibinfo{author}{\bibfnamefont{P.}~\bibnamefont{Frick}},
  \bibinfo{journal}{Phys. Rev. E} \textbf{\bibinfo{volume}{52}},
  \bibinfo{pages}{3719} (\bibinfo{year}{1995}).

\bibitem[{\citenamefont{Boffetta et~al.}(2000)\citenamefont{Boffetta, Celani,
  and Vergassola}}]{BCV00}
\bibinfo{author}{\bibfnamefont{G.}~\bibnamefont{Boffetta}},
  \bibinfo{author}{\bibfnamefont{A.}~\bibnamefont{Celani}}, \bibnamefont{and}
  \bibinfo{author}{\bibfnamefont{M.}~\bibnamefont{Vergassola}},
  \bibinfo{journal}{Phys. Rev. E} \textbf{\bibinfo{volume}{61}},
  \bibinfo{pages}{R29} (\bibinfo{year}{2000}).

\bibitem[{\citenamefont{Paret and Tabeling}(1998)}]{PT98}
\bibinfo{author}{\bibfnamefont{J.}~\bibnamefont{Paret}} \bibnamefont{and}
  \bibinfo{author}{\bibfnamefont{P.}~\bibnamefont{Tabeling}},
  \bibinfo{journal}{Phys. Fluids} \textbf{\bibinfo{volume}{10}},
  \bibinfo{pages}{3126} (\bibinfo{year}{1998}).

\bibitem[{\citenamefont{Honkonen}(1998)}]{H98}
\bibinfo{author}{\bibfnamefont{J.}~\bibnamefont{Honkonen}},
  \bibinfo{journal}{Phys. Rev. E} \textbf{\bibinfo{volume}{58}},
  \bibinfo{pages}{4532} (\bibinfo{year}{1998}).

\bibitem[{\citenamefont{Polyakov}(1993)}]{P93}
\bibinfo{author}{\bibfnamefont{A.~M.} \bibnamefont{Polyakov}},
  \bibinfo{journal}{Nucl. Phys. B} \textbf{\bibinfo{volume}{396}},
  \bibinfo{pages}{367} (\bibinfo{year}{1993}).

\bibitem[{\citenamefont{Forster et~al.}(1977)\citenamefont{Forster, Nelson,
and
  Stephen}}]{FNS77}
\bibinfo{author}{\bibfnamefont{D.}~\bibnamefont{Forster}},
  \bibinfo{author}{\bibfnamefont{D.~R.} \bibnamefont{Nelson}},
  \bibnamefont{and} \bibinfo{author}{\bibfnamefont{M.~J.}
  \bibnamefont{Stephen}}, \bibinfo{journal}{Phys. Rev. A}
  \textbf{\bibinfo{volume}{16}}, \bibinfo{pages}{732} (\bibinfo{year}{1977}).

\bibitem[{\citenamefont{DeDominicis and Martin}(1979)}]{DM79}
\bibinfo{author}{\bibfnamefont{C.}~\bibnamefont{DeDominicis}}
\bibnamefont{and}
  \bibinfo{author}{\bibfnamefont{P.~C.} \bibnamefont{Martin}},
  \bibinfo{journal}{Phys. Rev. A} \textbf{\bibinfo{volume}{19}},
  \bibinfo{pages}{419} (\bibinfo{year}{1979}).

\bibitem[{\citenamefont{Adzhemyan et~al.}(1999)\citenamefont{Adzhemyan,
  Antonov, and Vasiliev}}]{AAV99}
\bibinfo{author}{\bibfnamefont{L.~Ts.} \bibnamefont{Adzhemyan}},
  \bibinfo{author}{\bibfnamefont{N.~V.} \bibnamefont{Antonov}},
  \bibnamefont{and} \bibinfo{author}{\bibfnamefont{A.~N.}
  \bibnamefont{Vasiliev}}, \emph{\bibinfo{title}{The Field Theoretic
  Renormalization Group in Fully Developed Turbulence}}
  (\bibinfo{publisher}{Gordon and Breach}, \bibinfo{address}{Amsterdam},
  \bibinfo{year}{1999}).

\bibitem[{\citenamefont{Lesieur}(1997)}]{L97}
\bibinfo{author}{\bibfnamefont{M.}~\bibnamefont{Lesieur}},
  \emph{\bibinfo{title}{Turbulence in Fluids}} (\bibinfo{publisher}{Kluwer},
  \bibinfo{address}{Boston}, \bibinfo{year}{1997}), \bibinfo{edition}{3rd}
ed.

\bibitem[{\citenamefont{'t~Hooft and Veltman}(1972)}]{TV72}
\bibinfo{author}{\bibfnamefont{G.}~\bibnamefont{'t~Hooft}} \bibnamefont{and}
  \bibinfo{author}{\bibfnamefont{M.}~\bibnamefont{Veltman}},
  \bibinfo{journal}{Nucl. Phys. B} \textbf{\bibinfo{volume}{44}},
  \bibinfo{pages}{189} (\bibinfo{year}{1972}).

\bibitem[{\citenamefont{Martin et~al.}(1973)\citenamefont{Martin, Siggia, and
  Rose}}]{MSR73}
\bibinfo{author}{\bibfnamefont{P.~C.} \bibnamefont{Martin}},
  \bibinfo{author}{\bibfnamefont{E.~D.} \bibnamefont{Siggia}},
  \bibnamefont{and} \bibinfo{author}{\bibfnamefont{H.~A.}
\bibnamefont{Rose}},
  \bibinfo{journal}{Phys. Rev. A} \textbf{\bibinfo{volume}{8}},
  \bibinfo{pages}{423} (\bibinfo{year}{1973}).

\bibitem[{\citenamefont{Bausch et~al.}(1976)\citenamefont{Bausch, Janssen,
and
  Wagner}}]{BJW76}
\bibinfo{author}{\bibfnamefont{R.}~\bibnamefont{Bausch}},
  \bibinfo{author}{\bibfnamefont{H.~K.} \bibnamefont{Janssen}},
  \bibnamefont{and} \bibinfo{author}{\bibfnamefont{H.}~\bibnamefont{Wagner}},
  \bibinfo{journal}{Z. Phys. B} \textbf{\bibinfo{volume}{24}},
  \bibinfo{pages}{113} (\bibinfo{year}{1976}).

\bibitem[{\citenamefont{Phythian}(1977)}]{P77}
\bibinfo{author}{\bibfnamefont{R.}~\bibnamefont{Phythian}},
  \bibinfo{journal}{J. Phys. A} \textbf{\bibinfo{volume}{10}},
  \bibinfo{pages}{777} (\bibinfo{year}{1977}).

\bibitem[{\citenamefont{Adzhemyan et~al.}(1983)\citenamefont{Adzhemyan,
  Vasil'ev, and Pis'mak}}]{AVP83}
\bibinfo{author}{\bibfnamefont{L.~D.} \bibnamefont{Adzhemyan}},
  \bibinfo{author}{\bibfnamefont{A.~N.} \bibnamefont{Vasil'ev}},
  \bibnamefont{and} \bibinfo{author}{\bibfnamefont{Yu.~M.}
  \bibnamefont{Pis'mak}}, \bibinfo{journal}{Theor. Math. Phys. (USSR)}
  \textbf{\bibinfo{volume}{57}}, \bibinfo{pages}{1131}
(\bibinfo{year}{1983}).

\bibitem[{\citenamefont{'t~Hooft}(1973)}]{T73}
\bibinfo{author}{\bibfnamefont{G.}~\bibnamefont{'t~Hooft}},
  \bibinfo{journal}{Nucl. Phys. B} \textbf{\bibinfo{volume}{61}},
  \bibinfo{pages}{455} (\bibinfo{year}{1973}).

\bibitem[{\citenamefont{Deker and Haake}(1975)}]{DH75}
\bibinfo{author}{\bibfnamefont{U.}~\bibnamefont{Deker}} \bibnamefont{and}
  \bibinfo{author}{\bibfnamefont{F.}~\bibnamefont{Haake}},
  \bibinfo{journal}{Phys. Rev. A} \textbf{\bibinfo{volume}{11}},
  \bibinfo{pages}{2043} (\bibinfo{year}{1975}).

\bibitem[{\citenamefont{Frey and T{\"a}uber}(1994)}]{FT94}
\bibinfo{author}{\bibfnamefont{E.}~\bibnamefont{Frey}} \bibnamefont{and}
  \bibinfo{author}{\bibfnamefont{U.~C.} \bibnamefont{T{\"a}uber}},
  \bibinfo{journal}{Phys. Rev. E} \textbf{\bibinfo{volume}{50}},
  \bibinfo{pages}{1024} (\bibinfo{year}{1994}).

\bibitem[{\citenamefont{Adzhemyan et~al.}(2003)\citenamefont{Adzhemyan,
  Antonov, Kompaniets, and Vasil'ev}}]{AAKV03}
\bibinfo{author}{\bibfnamefont{L.~Ts.} \bibnamefont{Adzhemyan}},
  \bibinfo{author}{\bibfnamefont{N.~V.} \bibnamefont{Antonov}},
  \bibinfo{author}{\bibfnamefont{M.~V.} \bibnamefont{Kompaniets}},
  \bibnamefont{and} \bibinfo{author}{\bibfnamefont{A.~N.}
  \bibnamefont{Vasil'ev}}, \bibinfo{journal}{Int. J. Mod. Phys. B}
  \textbf{\bibinfo{volume}{17}}, \bibinfo{pages}{2137}
(\bibinfo{year}{2003}).

\bibitem[{\citenamefont{Wolfram}(1999)}]{W99}
\bibinfo{author}{\bibfnamefont{S.}~\bibnamefont{Wolfram}},
  \emph{\bibinfo{title}{The Mathematica Book}} (\bibinfo{publisher}{Cambridge
  University Press}, \bibinfo{address}{Cambridge}, \bibinfo{year}{1999}),
  \bibinfo{edition}{4th} ed.

\bibitem[{\citenamefont{Press et~al.}(1992)\citenamefont{Press, Teukolsky,
  Vetterling, and Flannery}}]{PTVF92}
\bibinfo{author}{\bibfnamefont{W.~H.} \bibnamefont{Press}},
  \bibinfo{author}{\bibfnamefont{S.~A.} \bibnamefont{Teukolsky}},
  \bibinfo{author}{\bibfnamefont{W.~T.} \bibnamefont{Vetterling}},
  \bibnamefont{and} \bibinfo{author}{\bibfnamefont{B.~P.}
  \bibnamefont{Flannery}}, \emph{\bibinfo{title}{Numerical Recipes in C: The
  Art of Scientific Computing}} (\bibinfo{publisher}{Cambridge University
  Press}, \bibinfo{address}{Cambridge}, \bibinfo{year}{1992}),
  \bibinfo{type}{section} \bibinfo{chapter}{7.8}, \bibinfo{edition}{2nd} ed.

\bibitem[{\citenamefont{Frisch and Sulem}(1984)}]{FS84}
\bibinfo{author}{\bibfnamefont{U.}~\bibnamefont{Frisch}} \bibnamefont{and}
  \bibinfo{author}{\bibfnamefont{P.~L.} \bibnamefont{Sulem}},
  \bibinfo{journal}{Phys. Fluids} \textbf{\bibinfo{volume}{27}},
  \bibinfo{pages}{1921} (\bibinfo{year}{1984}).

\bibitem[{\citenamefont{Danilov and Gurarie}(2001)}]{DG01}
\bibinfo{author}{\bibfnamefont{S.}~\bibnamefont{Danilov}} \bibnamefont{and}
  \bibinfo{author}{\bibfnamefont{D.}~\bibnamefont{Gurarie}},
  \bibinfo{journal}{Phys. Rev. E} \textbf{\bibinfo{volume}{63}},
  \bibinfo{pages}{020203(R)} (\bibinfo{year}{2001}).

\bibitem[{\citenamefont{Sommeria}(1986)}]{S86}
\bibinfo{author}{\bibfnamefont{J.}~\bibnamefont{Sommeria}},
  \bibinfo{journal}{J. Fluid Mech.} \textbf{\bibinfo{volume}{170}},
  \bibinfo{pages}{139} (\bibinfo{year}{1986}).

\bibitem[{\citenamefont{Maltrud and Vallis}(1991)}]{MV91}
\bibinfo{author}{\bibfnamefont{M.~E.} \bibnamefont{Maltrud}}
\bibnamefont{and}
  \bibinfo{author}{\bibfnamefont{G.~K.} \bibnamefont{Vallis}},
  \bibinfo{journal}{J. Fluid Mech.} \textbf{\bibinfo{volume}{228}},
  \bibinfo{pages}{321} (\bibinfo{year}{1991}).

\bibitem[{\citenamefont{Borue}(1994)}]{B94}
\bibinfo{author}{\bibfnamefont{V.}~\bibnamefont{Borue}},
  \bibinfo{journal}{Phys. Rev. Lett.} \textbf{\bibinfo{volume}{72}},
  \bibinfo{pages}{1475} (\bibinfo{year}{1994}).

\bibitem[{\citenamefont{Bernard}(1999)}]{B99}
\bibinfo{author}{\bibfnamefont{D.}~\bibnamefont{Bernard}},
  \bibinfo{journal}{Phys. Rev. E} \textbf{\bibinfo{volume}{60}},
  \bibinfo{pages}{6184} (\bibinfo{year}{1999}).

\bibitem[{\citenamefont{Dubos et~al.}(2001)\citenamefont{Dubos, Babiano,
Paret,
  and Tabeling}}]{DBPT01}
\bibinfo{author}{\bibfnamefont{T.}~\bibnamefont{Dubos}},
  \bibinfo{author}{\bibfnamefont{A.}~\bibnamefont{Babiano}},
  \bibinfo{author}{\bibfnamefont{J.}~\bibnamefont{Paret}}, \bibnamefont{and}
  \bibinfo{author}{\bibfnamefont{P.}~\bibnamefont{Tabeling}},
  \bibinfo{journal}{Phys. Rev. E} \textbf{\bibinfo{volume}{64}},
  \bibinfo{pages}{036302} (\bibinfo{year}{2001}).

\bibitem[{\citenamefont{Yakhot and She}(1988)}]{YS88}
\bibinfo{author}{\bibfnamefont{V.}~\bibnamefont{Yakhot}} \bibnamefont{and}
  \bibinfo{author}{\bibfnamefont{Z.-S.} \bibnamefont{She}},
  \bibinfo{journal}{Phys. Rev. Lett.} \textbf{\bibinfo{volume}{60}},
  \bibinfo{pages}{1840} (\bibinfo{year}{1988}).

\end{thebibliography}
\end{document}